\documentclass[journal=jacsat,manuscript=article,articletitle=true]{achemso}

\usepackage{chemformula} 
\usepackage[T1]{fontenc} 
\usepackage{amsmath}
\usepackage{amssymb}
\usepackage{titlesec}

\titleformat{\section}{\normalfont\Large\bfseries}{\thesection}{1em}{}

\author{Xiangyang Liu}
\affiliation{School of Materials Science and Engineering, University of Science and Technology of China, Shenyang 110016, China.}
\alsoaffiliation{Shenyang National Laboratory for Materials Science, Institute of Metal Research,Chinese Academy of Sciences, Shenyang 110016, China.}

\author{Junwen Lai}
\affiliation{School of Materials Science and Engineering, University of Science and Technology of China, Shenyang 110016, China.}
\alsoaffiliation{Shenyang National Laboratory for Materials Science, Institute of Metal Research,Chinese Academy of Sciences, Shenyang 110016, China.}

\author{Jie Zhan}
\affiliation{School of Materials Science and Engineering, University of Science and Technology of China, Shenyang 110016, China.}
\alsoaffiliation{Shenyang National Laboratory for Materials Science, Institute of Metal Research,Chinese Academy of Sciences, Shenyang 110016, China.}

\author{Tianye Yu}
\affiliation{Shenyang National Laboratory for Materials Science, Institute of Metal Research,Chinese Academy of Sciences, Shenyang 110016, China.}

\author{Peitao Liu}
\affiliation{School of Materials Science and Engineering, University of Science and Technology of China, Shenyang 110016, China.}
\alsoaffiliation{Shenyang National Laboratory for Materials Science, Institute of Metal Research,Chinese Academy of Sciences, Shenyang 110016, China.}

\author{Seiji Yunoki}
\affiliation{Computational Materials Science Research Team, RIKEN Center for Computational Science (R-CCS), Hyogo 650-0047, Japan}
\alsoaffiliation{Quantum Computational Science Research Team,RIKEN Center for Quantum Computing (RQC), Wako, Saitama 351-0198, Japan}
\alsoaffiliation{Computational Condensed Matter Physics Laboratory, RIKEN Cluster for Pioneering Research (CPR), Saitama 351-0198, Japan}
\alsoaffiliation{Computational Quantum Matter Research Team, RIKEN Center for Emergent Matter Science (CEMS), Saitama 351-0198, Japan}

\author{Xing-Qiu Chen}
\affiliation{School of Materials Science and Engineering, University of Science and Technology of China, Shenyang 110016, China.}
\alsoaffiliation{Shenyang National Laboratory for Materials Science, Institute of Metal Research,Chinese Academy of Sciences, Shenyang 110016, China.}
\email{xingqiu.chen@imr.ac.cn}

\author{Yan Sun}
\affiliation{School of Materials Science and Engineering, University of Science and Technology of China, Shenyang 110016, China.}
\alsoaffiliation{Shenyang National Laboratory for Materials Science, Institute of Metal Research,Chinese Academy of Sciences, Shenyang 110016, China.}
\email{sunyan@imr.ac.cn}
\title[An \textsf{achemso} demo]  {Nonlinear optical response in kagome lattice with inversion symmetry breaking}

\abbreviations{IR,NMR,UV}
\keywords{American Chemical Society, \LaTeX}

\begin{document}

\begin{abstract}
The kagome lattice is a fundamental model structure in 
condensed matter physics and materials science featuring 
symmetry-protected flat bands, saddle points, and Dirac
points. This structure has emerged as an ideal platform for exploring various 
quantum physics. By combining effective model analysis and first-principles calculations,
we propose that the synergy among inversion 
symmetry breaking, flat bands, and saddle point-related
van Hove singularities within the kagome lattice
holds significant potential for generating strong  
second-order nonlinear optical response. This property 
provides an inspiring insight into the practical application of the kagome-like materials,
which is helpful for a comprehensive understanding of kagome lattice-related physics. 
Moreover, this work offers an alternative approach for designing materials with strong a second-order nonlinear optical response.
\end{abstract}

Keywords: distorted kagome lattice, BPVE, shift current, flat bands, saddle points

\subsection*{INTRODUCTION}

The kagome lattice, characterized by the geometry of corner-sharing equilateral triangles,
is recognized as one of the most intriguing structures in condensed matter physics and materials science.
Its electronic structure is notable for the coexistence of the flat bands, saddle point related  van Hove singularities (VHSs), and Dirac points. 
Due to its special lattice and electronic structure, the kagome lattice serves as an ideal model platform for exploring the 
interplay of electron correlations, topological charges, frustrated magnetism, quantum phase transitions, and 
strong (or quantized) anomalous transport phenomena\cite{ghimire2020topology,kang2020dirac,neupert2022charge,yin2022topological,wang2023quantum}.

In addition to the aforementioned characteristics, the presence of flat bands and 
saddle points within the kagome lattice can result in a divergent density 
of states (DOS). The $\delta$-function in the formula of 
second-order nonlinear optical conductivity (see Eq. 6) indicates that this property establishes  
it as a natural platform for strong optical response, 
owing to the resulting large optical transition response\cite{kraut1979anomalous,von1981theory,campi1998formulation,orapunt2008optical,xiao2010berry,young2012first,tan2016enhancement,
gupta2018pursuit,qian2023shift,you2024versatile,tanaka2024nonlinear,laiuniversal,chen2024large}. 
However, a critical requirement for observing second-order nonlinear optical response is the breaking of inversion
symmetry. Fortunately, recent advancements in both experimental and theoretical research have provided several effective strategies
to achieve this. These strategies include the synthesis of compounds with distorted kagome lattice\cite{zhang2020single, wang2023synthesis, iandelli1973crystal, zhuravlev2010structure, duan2024observation, xie2024manipulation}, substrate-induced effects\cite{du2021engineering,zhou2024kagomerization,deng2024ferroelectricity,lee2023unconventional,li2024electronic},
phase transitions driven by external fields\cite{liu2024cascade, wang2022electronic, wu2024symmetry, li2019external, jiang2021flexo,zhang2023visualizing}, 
intercalation or defects\cite{zeng2020inversion,wu2024atomically,jo2022defect},
and heterostructure engineering or controlled layer stacking\cite{mu2023magnetic,gao2024bilayer}, among others. 
With these conditions, 
we propose a design principle aimed at generating strong
nonlinear optical response through the synergistic effect
of flat bands, and saddle point
related VHSs in the context of 
two-dimensional (2D) distorted kagome lattice. 
This work not only expands the scope of the kagome lattice research but also
provides a robust principle for designing materials with enhanced nonlinear optical functionalities.
By integrating theoretical insights with practical design principles, this study paves the way
for manipulating nonlinear optical response in the kagome lattice. 

\begin{figure*}[]
\centering
\includegraphics[width=0.95\textwidth]{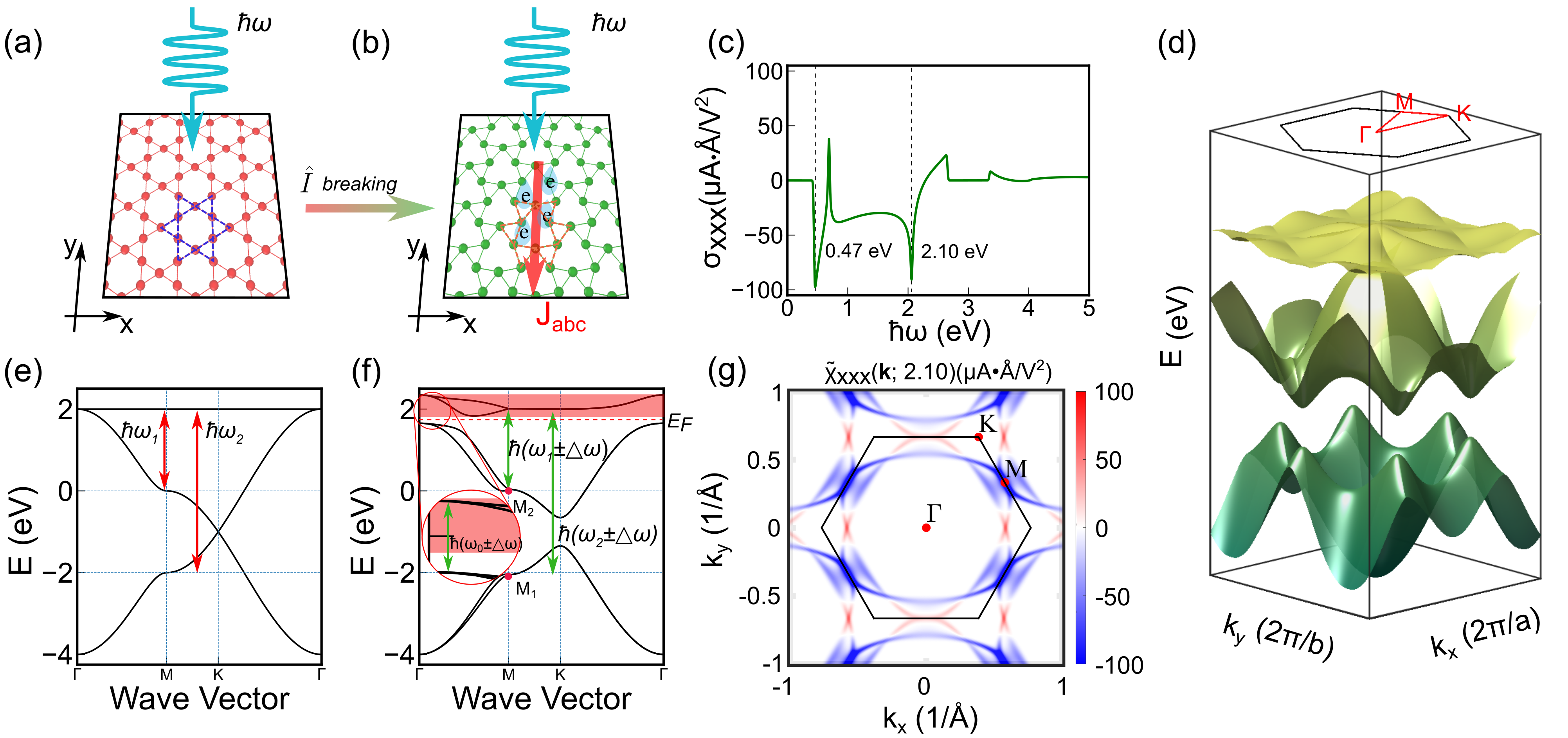}
\caption{Schematic of second-order nonlinear optical response and band structures of ideal and
distorted kagome lattices. (a) A net photocurrent is absent in an ideal kagome lattice with inversion symmetry.
(b) A non-zero net photocurrent is allowed in a distorted kagome lattice.
(c) Calculated second-order nonlinear optical conductivity for the lattice structure in (b).
(d) Energy dispersion for lattice structure (b) over the entir Brillouin zone.
(e) Energy dispersion for an ideal kagome lattice.
(f) Energy dispersion for a distorted kagome lattice after consideration of SOC and inversion symmetry breaking.
(g) Second-order nonlinear optical conductivity density distribution in momentum space with transition energy lying at $\hbar\omega$ = 2.1 eV.
The parameters $t_{1}$ = -1.0 eV, $t_{2}$ = 0.0 eV, and $\lambda_{SOC}$ = -0.2 eV were used.}
\label{model}
\end{figure*}

\subsection*{RESULTS AND DISCUSSION}
Starting from the conventional kagome lattice depicted in Figure 1a,
when considering the nearest electron hopping,
the kagome lattice hosts two saddle points at $M$ and 
one Dirac cone at $K$. Owing to the destructive 
interference of Bloch functions at the nearest-neighbor (NN) sites, the 
kinetic energy is fully quenched and the top band presents
as an infinite effective mass flat band over the entire Brillouin zone (BZ), 
as illustrated in Figure 1e\cite{ziman1971calculation,pommerenke1974bloch,beugeling2012topological,cook2017design,zhang2019kagome}. 
Generally, optical transitions from two 
saddle points to the flat bands, and even the transitions between flat bands 
and normal bands, can generate a strong optical transition response, which is believed to be a prerequisite for a large second-order nonlinear optical 
response.

Nevertheless, second-order nonlinear optical response
is forbidden by the inversion symmetry within the ideal
kagome lattice. To utilize the large optical transition response, 
we introduce a slight lattice distortion as an example   
to break the inversion symmetry while preserving the shape of the
saddle points and flat bands. To achieve this, one can rotate the adjacent triangles in 
opposite directions, resulting in 
a Reuleaux-triangle-shaped lattice configuration, as depicted in Figure 1b.
The rotation angle can vary from 0$^\circ$ to 30$^\circ$, corresponding to a kagome lattice and a coloring-triangular lattice, 
respectively (see Figure S1a-c). 
Correspondingly, a six-band effective Hamiltonian is constructed in momentum space 
is given by\cite{beugeling2012topological,zhang2019kagome}.
\begin{equation}
  \mathcal{H}=\mathcal{H}^{\mathrm{NN}}\otimes\mathbb{I}_{2\times2} +\mathcal{H}^{\mathrm{NNN}}\otimes\mathbb{I}_{2\times2}+\mathcal{H}^{\mathrm{SOC}}\otimes\sigma_{z},
\end{equation}
where the first two terms $\mathcal{H}^{\mathrm{NN}}$ and $\mathcal{H}^{\mathrm{NNN}}$ describe the NN and next NN (NNN) hoppings, respectively.
Spin-orbit coupling (SOC) is included here to open the gap between the fourth and fifth bands, which is described as the third term $\mathcal{H}^{\mathrm{SOC}}$. The 3 $\times$ 3 matrices $\mathcal{H}^{\mathrm{NN}}$, $\mathcal{H}^{\mathrm{NNN}}$, and 
$\mathcal{H}^{\mathrm{SOC}}$ are given by
\begin{equation}
  \mathcal{H}^{\mathrm{NN}}=t{_1}\begin{pmatrix}0&c_1&c_3\\c_1^{*}&0&c_2\\c_3^{*}&c_2^{*}&0\end{pmatrix},  
\end{equation}
\begin{equation}
 \mathcal{H}^{\mathrm{NNN}}=t{_2}\begin{pmatrix}0&b_1&b_3\\b_1^{*}&0&b_2\\b_3^{*}&b_2^{*}&0\end{pmatrix}, 
\end{equation}
and
\begin{equation}
  \mathcal{H}^{\mathrm{SOC}}=i\lambda_{\mathrm{SOC}}\begin{pmatrix}0&b_1&-b_3\\-b_1^{*}&0&b_2\\b_3^{*}&-b_2^{*}&0\end{pmatrix},
\end{equation}
where, $c_{1} = e^{i\vec{k}\cdot\frac{\vec{v}_{1}-\vec{v}_{3}}{3}}+e^{i\vec{k}\cdot\frac{\vec{v}_{3}-\vec{v}_{2}}{3}}$, 
$c_{2} = e^{i\vec{k}\cdot\frac{\vec{v}_{2}-\vec{v}_{1}}{3}}+e^{i\vec{k}\cdot\frac{\vec{v}_{1}-\vec{v}_{3}}{3}}$,
$c_{3} = e^{-i\vec{k}\cdot\frac{\vec{v}_{3}-\vec{v}_{2}}{3}}+e^{-i\vec{k}\cdot\frac{\vec{v}_{2}-\vec{v}_{1}}{3}}$,
$b_{1} = e^{i\vec{k}\cdot\frac{\vec{v}_{2}-\vec{v}_{1}}{3}}$,
$b_{2} = e^{i\vec{k}\cdot\frac{\vec{v}_{3}-\vec{v}_{2}}{3}}$,
and $b_{3} = e^{i\vec{k}\cdot\frac{\vec{v}_{3}-\vec{v}_{1}}{3}}$.
Here, $\vec{v}_{1}$ and $\vec{v}_{2}$ represent the basis 
vectors of the distorted kagome primitive cell, with $\vec{v}_{3}$ 
defined as $-(\vec{v}_{1} + \vec{v}_{2})$, while the $t_{i}$ is the hopping integral (see Figure S2a). 

With the given hopping parameters and considering the influence of SOC, 
the lattice distortion alters the band structure from Figure 1d,f.
One can see that although the Dirac cone
at $K$ is broken by opening a band gap, two saddle
points at $M$ persist, and the wrinkling magnitude $\hbar\Delta\omega$ of flat bands is small. 
Furthermore, a full band gap emerges between the fourth and fifth bands, and local flat regions form in the third and the fourth bands around the $\Gamma$ point.
As a result, the large optical transition response is maintained after the lattice distortion with 
inversion symmetry breaking, and a strong second-order nonlinear optical response is expected (see Figure S2b-c).
 The bulk photovoltaic effect (BPVE) can generate a steady photocurrent with photovoltage above 
the band gap. According to the current understanding, the shift current is believed to be one of the 
major origins of the BPVE. Therefore we will mainly focus on the shift current.\cite{young2012first,cook2017design,ibanez2018ab,zhang2019enhanced,sauer2023shift2,liang2023strong}.

As depicted in Figure 1c, with the chemical potential set between the
fourth and fifth bands, two distinct peaks of second-order nonlinear optical conductivity
emerge near the photon energies of 0.47 eV and 2.10 eV.
After analyzing the band structure in Figure 1f, one can observe that the peak near 0.47 eV ($\hbar (\omega_0 \pm \Delta\omega)$)
corresponds to the optical transition between the fourth band and the flat bands
near the $\Gamma$ point, which is indicated by the arrow in the inset. Similarly,
the second peak near 2.10 eV ($ \hbar (\omega_1 \pm \Delta\omega)$) aligns with the optical transition from the
saddle point $M_2$ to the flat bands. This speculation is further supported
by the detailed analysis of the DOS and transition energy isosurfaces in Figure S3.
Since a large optical transition response is only a prerequisite but not a sufficient condition
for the strong nonlinear optical response, a large shift current corresponding
to the optical transition between the saddle point $M_1$ and the flat bands does not appear
in Figure 1c.

\begin{figure}[]
\includegraphics[width=0.7\textwidth]{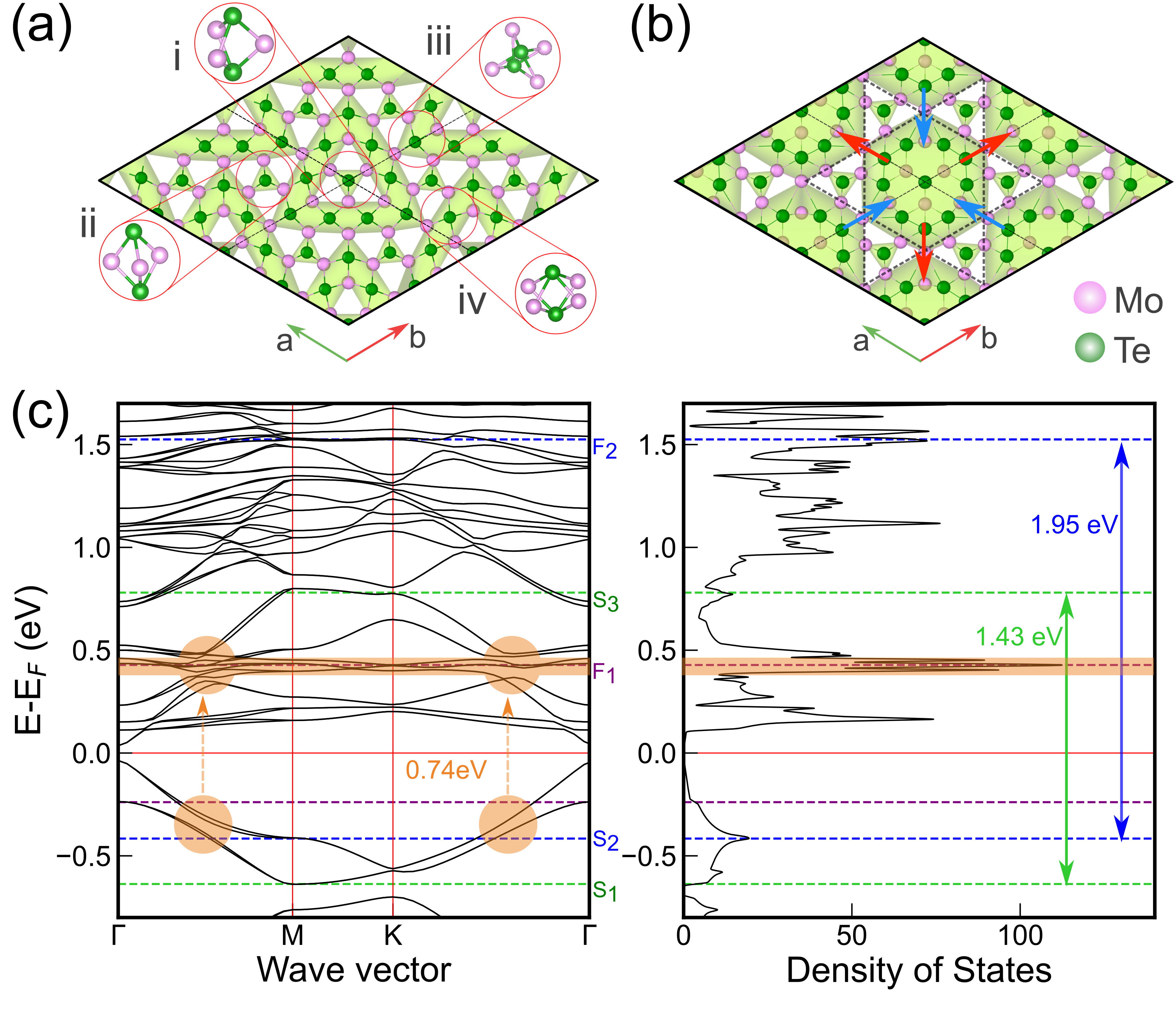}
\caption{Geometry schematic and electronic structures of Mo$_5$Te$_8$. 
(a) Crystal structure of $2 \times 2$ supercell of Mo$_5$Te$_8$.
(b) The parent ideal kagome structure for Mo$_5$Te$_8$ can be derived from (a) by 
moving the atoms along the arrows.
(c) Energy dispersion and density of states of Mo$_5$Te$_8$.}
\label{real_lattice}
\end{figure}

To further confirm the relationship between the strong shift current and the two special 
electronic band structures, we analyzed the 
density distribution of the second-order nonlinear optical
conductivity. Taking the photon energy of 2.10 eV
as an example, one can find that the biggest contribution of hot spots comes from the area near the $M$ point in Figure 1g, which
reflects the main contribution from the optical transition 
between saddle point to flat bands. 
Similarly, at 0.47 eV, as illustrated in Figure S3b-c, the shift current density primarily originates from
six elliptical regions around the $\Gamma$ point, 
corresponding to the optical transition from the top of the fourth band near the $\Gamma$ 
point to the bottom of the flat bands.
Therefore, the interplay between inversion symmetry 
breaking and two distinctive band structures within the
kagome lattice holds significant potential for generating
a robust second-order nonlinear optical response. 

\begin{figure}[]
\centering
\includegraphics[width=0.7\textwidth]{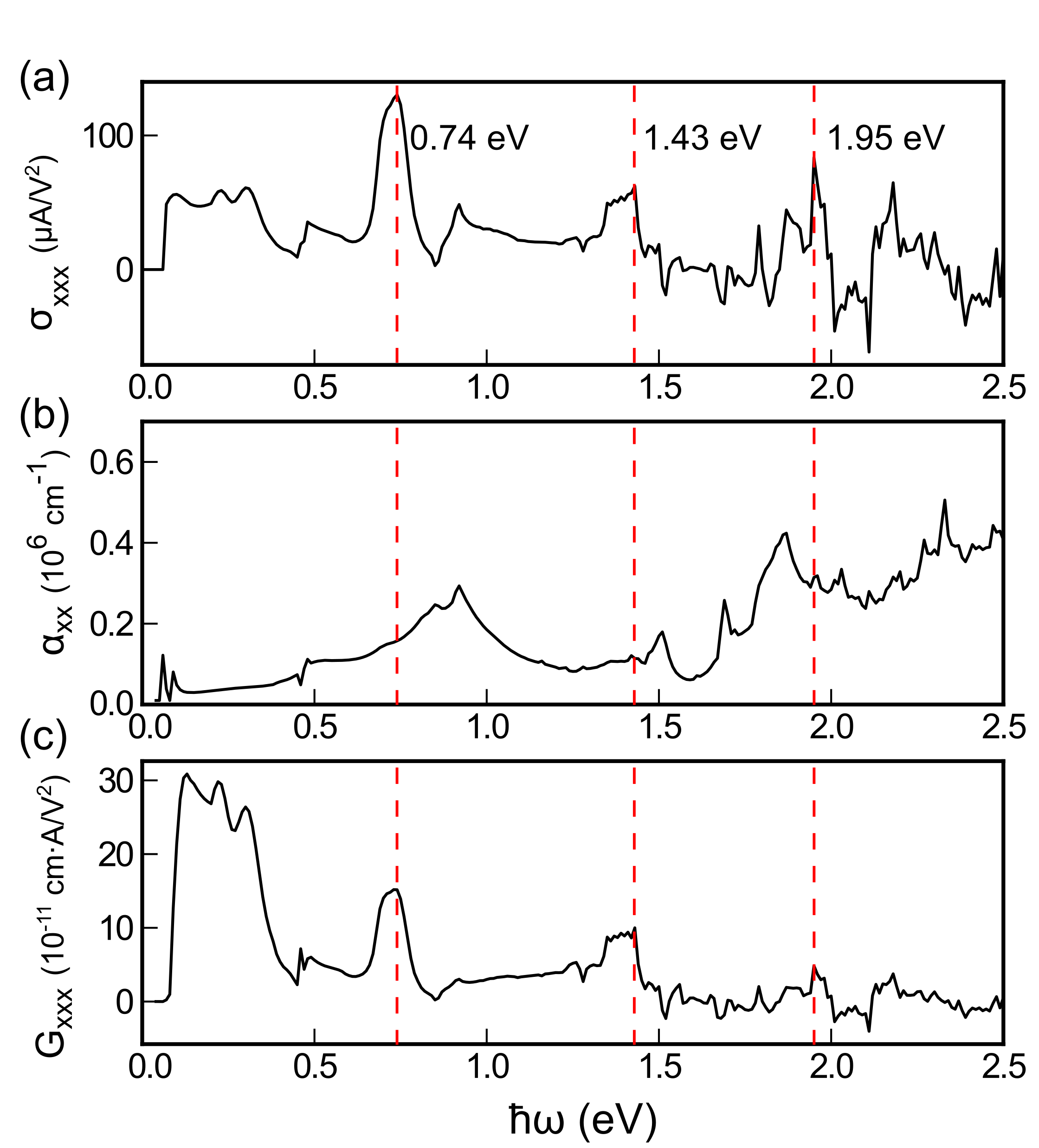}
\caption{Optical responses in monolayer Mo$_5$Te$_8$. (a) Shift current spectrum of $\sigma_{xxx}$ tensor component, (b) absorption
coefficient, and (c) Glass coefficient.}
\label{sc}
\end{figure}

Drawing on insights from the model analysis, we now search for materials with distorted kagome lattices. 
Through symmetry analysis, the distorted structure in Figure 1b adopts the layer group of $P\bar{6}2m$(No. 79). 
Guided by this information, we identify  
a synthesized material formula Mo$_5$Te$_8$, which meets the criteria mentioned above\cite{zhang2020single,lei2023electronic}. 
Its primitive cell consists of 15 Mo and 24 Te atoms. 
As illustrated in Figure 2a, a pristine 1H-MoTe$_2$ unit is enveloped by 
three uniform mirror twin boundaries (MTBs), forming a Reuleaux-triangle-shaped structural unit.
When the corners of these triangles connect with the midpoints of adjacent triangles, 
the overall pattern of Mo$_5$Te$_8$ emerges. 
This arrangement results in four distinct local atomic environments centered around Te atoms, 
each highlighted by a red circle in Figure 2a. 
The MTB regions are denser, primarily constructed of type-$iii$ and type-$iv$ atoms, 
while type-$i$ and type-$ii$ atoms are embedded within the framework of MTBs.
Based on the highlighted MTBs region in Figure 2a and disregarding the details 
enclosed by the MTBs, one can identify that Mo$_5$Te$_8$ 
adopts a distorted kagome lattice structure accompanied by the 
inversion symmetry breaking. The detailed relationship between 
them is illustrated in Figures S1 and S7a. Naturally, Mo$_5$Te$_8$ possesses 
a standard kagome lattice counterpart,
achievable through folding it along the six distinct directions depicted by the arrows in the Figure 2b.

The electronic band structure and DOS are depicted 
in Figure 2c. Since there are 
four types of Mo atoms in the primitive cell, and only one type
forms the distorted kagome lattice, the 
hybridization between bands from the distorted 
kagome lattice and others is relatively strong.
Fortunately, the features of saddle points and flat bands
from Mo can still be easily recognized in the electronic
band structure, as shown in Figure 2c and Figure S5-S7. 
Additionally, the physics near the Fermi energy is dominated by 
the kagome-type band structures, which are energetically separated from other occupied 
bands due to the presence of 
MTBs within its geometry. 
Furthermore, the band structure could be tuned through controlling 
the size of MTBs\cite{tan2016shift,bihari2019large,tan2019upper,dai2024kagome,hong2017inversion,liu2014dense}, 
making Mo$_5$Te$_8$ a compelling candidate for investigating the role of the distorted kagome
lattice in the bulk photovoltaic effect. 
Moreover, atomically thin two-dimensional materials hold significant promise as building blocks for next-generation shift current devices\cite{rangel2017large,chan2021giant,xu2021colossal,xiao2022non,jin2024peculiar}. Therefore, Mo$_5$Te$_8$ is a potential material candidate for the study of
shift current. 
In detail, there are four groups of flat bands near the energies of 0.16 eV, 
0.43 eV, 1.10 eV, and 1.52 eV, which correspond to four prominent peaks of DOS. Moreover, four saddle 
points align with high-symmetry points $M$ at energies approximately at -0.63 eV,
-0.41 eV, 0.5 eV, and 0.8 eV. These points also appear as VHSs in the DOS.
Therefore, Mo$_5$Te$_8$ inherits the features of kagome lattice, which could generate a large optical transition response and 
hold significant potential to produce strong second-order nonlinear optical response.

Following the crystal structure of Mo$_5$Te$_8$ adopts
the layer group of $P\bar{6}2m$(No. 79), the second-order nonlinear optical
conductivity tensor follows the shape of 
\begin{equation}
  \begin{bmatrix}\chi_{xxx}&-\chi_{xxx}&0&0&0&0&0&0&0\\0&0&0&0&0&0&0&-\chi_{xxx}&-\chi_{xxx}\\0&0&0&0&0&0&0&0&0\end{bmatrix}
\end{equation}
with only one independent component $\chi_{xxx}$, where the row index corresponds to \{x, y, z\} and the column index 
corresponds to \{xx, yy, zz, yz, zy, xz, zx, xy, yx\}. The results from numerical calculations 
are in full agreement with the symmetry analysis. The peak value of $\sigma_{xxx}$ can reach up to 
approximately 100 $\mu$A/V$^2$ at the optical transition energy $\sim$0.74 eV,
which is a relative strong response signal\cite{young2012first2,young2012first,brehm2014first,wang2017first,cook2017design,rangel2017large,ibanez2018ab,zhang2019switchable,sotome2019spectral,chan2019exciton,
fei2020shift,schankler2021large,sauer2023shift2,jin2024peculiar}, as illustrated in Figure 3a. In addition, two notable peaks
near $\hbar\omega\sim$1.43 eV and $\hbar\omega\sim$1.95 eV 
are also exceed 50 $\mu$A/V$^2$. 
As expected, through the optical transition energy matching, the photon energies 
of the three peaks above 50 $\mu$A/V$^2$ are closely related to the optical transition involving flat bands 
and saddle points, as labeled in Figure 2c and Figure S6. 
Hence, it is clear that flat bands or saddle points play key roles in  
the second-order optical response of the distorted kagome lattice. 
\begin{figure*}[]
\centering
\includegraphics[width=0.85\textwidth]{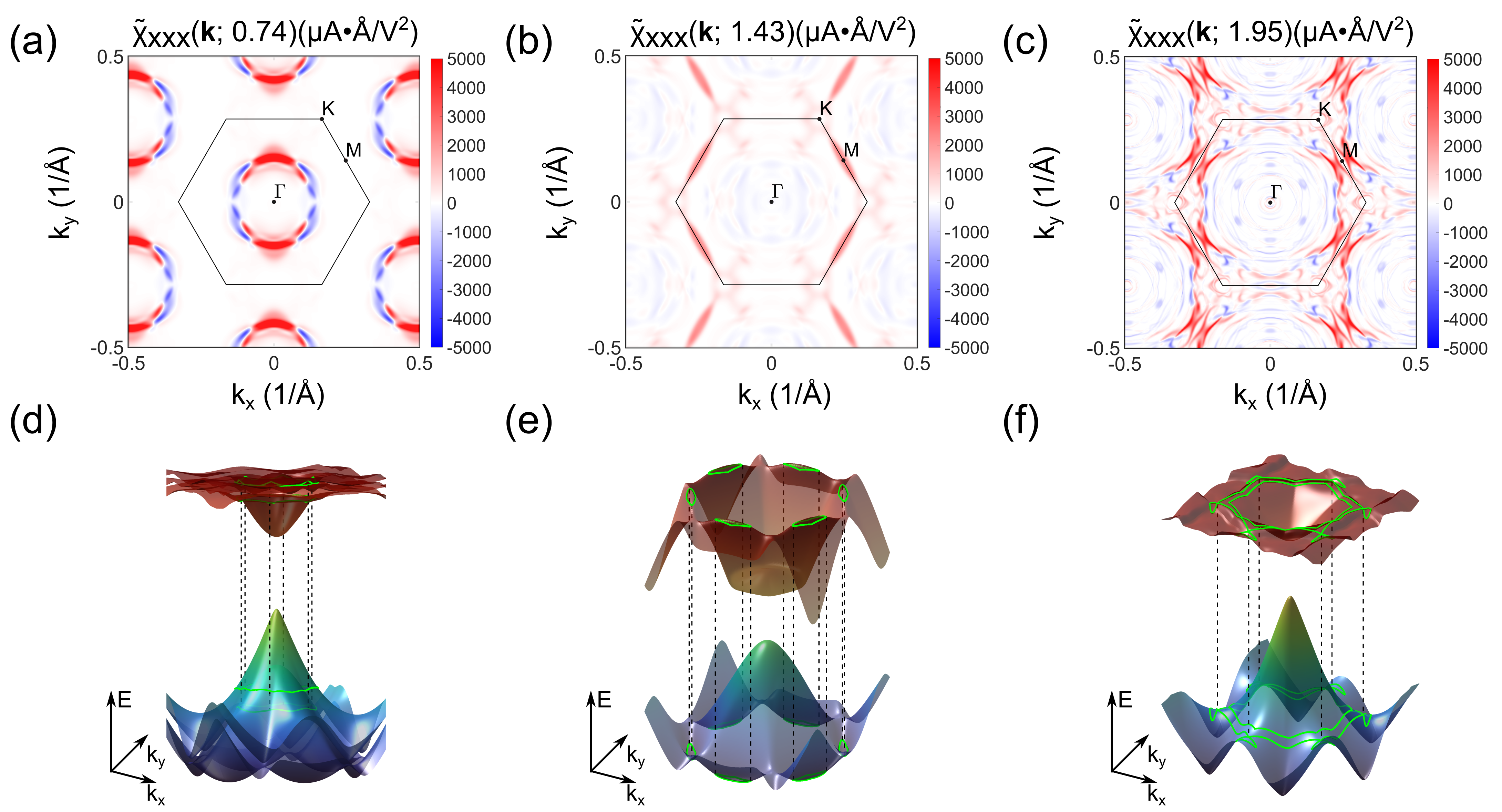}
\caption{Relation between the second-order nonlinear optical conductivity density distribution 
of $\tilde\chi_{xxx}(\textbf{k}; \omega)$ and electronic band structures. 
(a-c) The distribution of  $\tilde\chi_{xxx}(\textbf{k}; \omega)$ with photon
energy at 0.74 eV, 1.43 eV and 1.95 eV, respectively.
(d-f) Light excitation between occupied bands and non-occupied
bands corresponding to (a-c).}
\label{distribution}
\end{figure*}

To further validate the relationship
between the peak values of the second-order nonlinear optical 
conductivity and the band structure characteristic of the kagome lattice in Mo$_5$Te$_8$, 
we depicted the conductivity density distribution 
by fixing the photon energies at the three peaks of the shift current. 
As presented in Figure 2c and Figure 4a,d,
the first peak of the second-order nonlinear optical 
conductivity is dominated by optical transitions 
from a circular region near the high-symmetry point $\Gamma$ to the flat bands group F$_1$. 
In details, a bunch of flat bands near 0.45 eV collectively contribute to 
the shift current response at $\hbar\omega$ $\sim$0.74 eV and thus a 
relatively broader peak is present.
This suggests that in real materials, the peak magnitude and the broadening of the optical conductivity can also be
enhanced by the contribution from multiple flat bands. 
The peaks near $\hbar\omega$ $\sim$1.43 eV and $\hbar\omega$  $\sim$1.95 eV 
are mainly contributed by the optical transitions
from the saddle point S$_1$ to the saddle point
S$_3$ and the saddle point S$_2$ to the flat bands F$_2$, as shown in Figures 2c and Figure 4b-c,e-f, respectively. 

In the design of actual devices, optical absorption properties play a 
crucial role in governing the intensity of the second-order nonlinear optical response. 
To account for light attenuation, the intensity of the second-order nonlinear 
optical response is often evaluated using the Glass coefficient, defined as 
$G_{abb}=\frac{\sigma_{abb}(\omega)}{\alpha_{bb}(\omega)}$, where $\sigma_{abb}$
is the second-order nonlinear optical conducitivity and $\alpha_{bb}(\omega)$ 
is the absorption coefficient\cite{glass1995high,lalitha2007electronic,young2012first2,rangel2017large,ma2021topology,brehm2014first}.
As illustrated in the Figure 3b-c, the absorption coefficient and Glass coefficient 
of monolayer Mo$_5$Te$_8$ are provided. Notably, the Glass coefficient 
is on the order of $\sim 10^{-10}$ cm$\cdot$AV$^{-2}$, which is comparable to that of monolayer monochalcogenides\cite{rangel2017large}.
Furthermore, when considering the temperature effect 
on the shift current response (see Figure S11), one can find that 
even at 300 K, only the response intensity within the frequency range of 
0.3 eV is weaken, while the peak values remain largely unchanged when they lie outside this range\cite{yang2017divergent}. 
Additionally, to reflect intrinsic material losses such as disorder 
or inelastic scattering processes, we replace the delta function by a Lorentzians 
in Eq. 6\cite{chaubey1986transverse}. As shown in Figure S12, the intensity of the 
second-order nonlinear optical response decrease as the broadening width increases.
This highlights the importance of intrinsic material losses in the design
of BPVE devices\cite{sotome2019spectral}.

To generalize our understanding of this mechanism, we extended the compound Mo$_5$Te$_8$
by substituting its elements with others from the same 
group in the periodic table. Drawing inspiration from the strong shift current predicted in 
Mo$_5$Te$_8$, five additional counterpart compounds - W$_5$S$_8$, Mo$_5$Se$_8$, 
W$_5$Se$_8$, Mo$_5$S$_8$, and W$_5$Te$_8$, were also studied. 
Although only Mo$_5$Te$_8$ has been experimentally synthesized thus far, 
we conducted calculations to evaluate the stability for the other five compounds from thermodynamically and dynamically. 
The results indicate that the formation energies of these compounds are
negative and no imaginary phonon modes were identified\cite{togo2015first,phonolammps} (see Figure S4). Hence, all 
these five compounds are potentially stable. The calculated second-order nonlinear
optical conductivity spectra are shown in Figure S10b-f,
where four out of five compounds exhibit peak values exceeding 100 $\mu$A/V$^2$, 
and all feature a sizeable photon energy transfer window with a value above 50 $\mu$A/V$^2$. This suggests a general principle 
that the kagome lattice with inversion symmetry breaking can yield strong 
second-order nonlinear optical responses. Additionally, the relationship between transition energy, electronic properties, 
and the second-order nonlinear optical conductivity density distribution in momentum space was analyzed for each counterpart compound, 
as shown in Figure S8 and Figure S9. 
Furthermore, the effects of temperature and intrinsic material loss on the second-order nonlinear 
optical response are provided in Figure S11 and Figure S12, respectively.
Notably, all observed cases follow the principles outlined above. Therefore, both the 
location of optical transition and the density distribution of the 
second-order nonlinear optical conductivity in momentum space confirm that shift current response in the
distorted kagome lattice strongly depends on lattice-induced saddle points and flat bands.

\subsection*{CONCLUSION}
In summary, through effective model analysis,
we demonstrated that the interplay between
inversion symmetry breaking and special
electronic structures of saddle points and 
flat bands within the kagome lattice is an effective approach to generate
strong nonlinear optical response. Following this
guiding principle, we have identified real
materials like Mo$_5$Te$_8$ with a distorted
kagome lattice that not only breaks inversion symmetry but also preserves
flat bands and saddle points. The peak value of shift current in Mo$_5$Te$_8$ and its 
counterpart compounds can reach up to approximately 100 $\mu$A/V$^2$. 
This study provides another insight 
into kagome lattice-related physics and materials from the perspective of the nonlinear quantum response. 
While our focus in this study was primarily on examining inversion symmetry breaking in the kagome lattice by 
rotating the triangles, we mentioned above that there are other methods can also break its inversion
symmetry. We believe some of these alternative methods could yield similar effects in generating 
strong second-order nonlinear optical responses. This work is expected to attract further research
efforts to expand the understanding of kagome lattice in the context of the second-order nonlinear optical response.

\subsection*{METHODS}
\subsubsection*{DFT calculation details}
All density functional theory (DFT) calculations presented in this work were conducted using
the Vienna ab initio simulation package (VASP)\cite{kresse1996efficient,kresse1996efficiency} within the Perdew-Burke-Ernzerhof (PBE)\cite{ernzerhof1999assessment}
 generalized gradient approximation (GGA) for the exchange-correlation functional.
 The hybrid functional (HSE06)\cite{krukau2006influence} was also employed to obtain a more accurate band structures.
 The projector augmented wave method was used, with a plane-wave energy cutoff of 300 eV and the
 \textbf{k}-point mesh with the reciprocal space resolution of 2$\pi$ $\times$ 0.02 $\mathring{A}^{-1}$.
For geometry optimization, the convergence threshold for atomic force and energy is 0.01 eV/$\mathring{A}$ 
and 1 $\times$ $10^{-5}$ eV, respectively. To reduce the interaction effects between two adjacent layers, 
a vacuum layer with a width of around 15 $\mathring{A}$ was introduced in the non-periodic direction.
 
\subsubsection*{Calculation of the second-order nonlinear optical response}
The shift current density is considered as a second-order nonlinear optical response to the electromagnetic field E of frequency $\omega$,
\begin{equation}
J_a = \sigma_{abc}(0;\omega,-\omega)E_b(\omega)E_c(-\omega)  
\end{equation}
where the third-rank tensor $\sigma_{abc}(0; \omega, -\omega)$ is the shift current response tensor with Cartesian index $a$, $b$, and $c$. 
Index $a$ denotes the direction of the generated DC current, while $b$ and $c$ correspond to the polarization directions of incident light. 
\begin{equation}
\sigma_{abc}(0;\omega,-\omega)=\sum_{k}\tilde{\chi }_{abc}(\textbf{k}; \omega),    
\end{equation}

\begin{equation}
\begin{aligned}
\tilde{\chi }_{abc}(\textbf{k};\omega) = \sum_{m,n}Af_{nm}^{\textbf{k}}(r_{\textbf{k}, mn}^{b}r_{\textbf{k}, nm}^{c;a} + r_{\textbf{k}, mn}^{c}r_{\textbf{k}, nm}^{b;a})
\delta(\hbar\omega - E_{\textbf{k},mn}),
\end{aligned}
\end{equation}

\begin{equation}
\begin{aligned}
r_{\textbf{k}, nm}^{a} = i\langle n | \partial_{k_a} |m \rangle = \frac{\upsilon_{\textbf{k}, nm}^{a}}{i\omega_{\textbf{k,nm}}},
\end{aligned}
\end{equation}

\begin{equation}
\begin{aligned}
r_{\textbf{k}, nm}^{a;b}
= \partial_{k_b}r_{\textbf{k},nm}^{a}+(\langle n | \partial_{k_b} |n \rangle -\langle m | \partial_{k_b} |m \rangle )r_{\textbf{k},nm}^{a}\\
=\frac{i}{\omega_{\textbf{k},nm}}[\upsilon_{\textbf{k},nm}^{a}\varDelta_{\textbf{k},nm}^{a} + \upsilon_{\textbf{k},nm}^{b}\varDelta_{\textbf{k},nm}^{b} - \omega_{\textbf{k},nm}^{ab} \\
+ \sum_{p\neq n,m}(\frac{\upsilon_{\textbf{k},np}^{a}\upsilon_{\textbf{k},pm}^{b}}{\omega_{\textbf{k},pm}} - \frac{\upsilon_{\textbf{k},np}^{b}\upsilon_{\textbf{k},pm}^{a}}{\omega_{\textbf{k},np}})],
\end{aligned}
\end{equation}

\noindent
where n and m are band indexes, and \textbf{k} represents the wavevector of the Bloch wave function. 
$A = \frac{ie^{3}\pi}{\hbar V}$ is the coefficient of shift current. 
$f_{nm}^{k} = f_{n}^{k} - f_{m}^{k}$
 signifies the difference in the Fermi-Dirac occupation
numbers between band n and m, while $E_{\textbf{k}, nm} = E_{\textbf{k},n} - E_{\textbf{k},m}$ 
is the energy difference between band $m$ and $n$. 
$\upsilon_{\textbf{k},nm}^{a} = \frac{1}{\hbar} \langle n|\partial_{a} \hat{H}|m \rangle$ is the velocity matrix.
$ \varDelta_{\textbf{k}, nm}^{a} = \upsilon_{\textbf{k}, nn}^{a} - \upsilon_{\textbf{k}, mm}^{a} $ is the difference in Fermi velocity between the band $m$ 
and $n$, and $\omega_{\textbf{k}, nm}^{ab} = \frac{1}{\hbar} \langle n | \partial_{ab}^{2} \hat{H}|m \rangle$\cite{qian2023shift,kraut1979anomalous,
von1981theory,young2012first,ibanez2018ab,xiao2010berry}.

Maximum localized Wannier function (MLWF) method was carried out to construct 
 the tight binding models of compounds investigated in this study. This was 
 conducted by employing the Wannier90 package\cite{mostofi2008wannier90} based on the ground state results from VASP.
 
Shift current reported in this study was calculated using the Wannier90 package with a broadening parameter as 10 meV. 
To incorporate the effects of temperature and intrinsic material losses, an in-house code is employed.

The absorption coefficient is calculated with the following equation\cite{lalitha2007electronic}

\begin{equation}
\begin{aligned}
\alpha(\omega) = \sqrt{2}\omega[\sqrt{\varepsilon_1^2(\omega)+\varepsilon_2^2(\omega)} - \varepsilon_1 (\omega)^{1/2}],
\end{aligned}
\end{equation}

\noindent
where $\varepsilon_2(\omega)$ and $\varepsilon_2(\omega)$ are the real and imaginary parts of the dielectric function, respectively.
The optical absorption coefficient reported in this study is calculated based on Wannier function interpolation of a hybrid density 
functional theory band structure\cite{kim2021ab}.

\subsubsection*{Calculation of phonon dispersion}
The phonon dispersion curves were computed using the Deep potential (DP) and phonolammps package\cite{togo2015first,phonolammps}. 
As shown in the Figure S4, the phonon dispersion results for Mo$_5$Te$_8$ obtained with DP potential agree well with that calculated using VASP. 
Thus, the scheme used in this study is reliable for phonon dispersion calculations.

\subsection*{ASSOCIATED CONTENT}
\textbf{Supporting Information}
The supporting Information is available free of charge at \url{https://xxx.xxx}.

Comparison between the kagome lattice and the Mo$_5$Te$_8$ lattice (Figure S1);
Variation of electronic and optical properties with hopping parameters in the distorted kagome lattice (Figure S2);
Electronic properties, second-order nonlinear optical conductivity density distribution, 
and transition energy isosurfaces of the distorted kagome lattice model (Figure S3);
Phonon dispersion relations of monolayer Mo$_5$Te$_8$, W$_5$S$_8$, Mo$_5$Se$_8$, W$_5$Se$_8$, Mo$_5$S$_8$, and W$_5$Te$_8$ (Figure S4);
Weighted projected band structures of monolayer Mo$_5$Te$_8$ (Figure S5);
Band structure of monolayer Mo$_5$Te$_8$ highlighting kagome-like bands near the Fermi level (Figure S6);
Lattice decomposition of monolayer Mo$_5$Te$_8$ and projected band structure contributions from different Mo atoms (Figure S7);
Band structures and density of states (DOS) of W$_5$S$_8$, Mo$_5$Se$_8$, W$_5$Se$_8$, Mo$_5$S$_8$, and W$_5$Te$_8$ (Figure S8);
Second order nonlinear optical conductivity density distribution in momentum space (Figure S9);
Shift current spectrum, absorption coefficient, and Glass coefficient of monolayer W$_5$S$_8$, Mo$_5$Se$_8$, W$_5$Se$_8$, Mo$_5$S$_8$ , and W$_5$Te$_8$ (Figure S10);
Temperature-dependent shift current spectrum of W$_5$S$_8$, Mo$_5$Se$_8$, W$_5$Se$_8$, Mo$_5$S$_8$, and W$_5$Te$_8$ (Figure S11);
Effect of Dirac-Delta function forms and broadening parameters on the shift current spectrum of monolayer Mo$_5$Te$_8$, W$_5$S$_8$, Mo$_5$Se$_8$, W$_5$Se$_8$, 
Mo$_5$S$_8$, and W$_5$Te$_8$ (Figure S12)

The tight-binding Hamiltonians are available at \url{https://github.com/Dustglaxy/TB_files/blob/main/hr_files.tar}

\subsection*{AUTHOR INFORMATION}

\textbf{Author Contributions}
Y.S. and X.C. conceived the idea and designed the research.
X.L., J.L. and Y.S. performed the research and analyzed the data.
All authors contributed to the discussion and writing of the manuscript.
X.L. and J.L. contributed equally to this work.

\setlength{\parindent}{0pt}
\textbf{Notes}
The authors declare no competing financial interest.
\begin{acknowledgement}
The authors would like to thank ljaz Shahid and Xiliang Gong for carefully reading the manuscript. 
This work was supported by the Liao Ning Revitalization Talents Program (Grant No. XLYC2203080), National
Natural Science Foundation of China (Grants No. 52271016, No. 52188101, and No.52422112), National Key Research and Develop-
ment Program of China (Grant No. 2021YFB3501503), the National Chinese Academy of Sciences Project for Young
Scientists in Basic Research (Grant No. YSBR-109), and the Special Projects of the Central Government in Guidance
of Local Science and Technology Development (Grant No.2024010859-JH6/1006).
Part of the calculations were carried out on the ORISE Supercomputer (Grant No. DFZX202319).
\end{acknowledgement}

\bibliography{references}

\providecommand{\latin}[1]{#1}
\makeatletter
\providecommand{\doi}
  {\begingroup\let\do\@makeother\dospecials
  \catcode`\{=1 \catcode`\}=2 \doi@aux}
\providecommand{\doi@aux}[1]{\endgroup\texttt{#1}}
\makeatother
\providecommand*\mcitethebibliography{\thebibliography}
\csname @ifundefined\endcsname{endmcitethebibliography}
  {\let\endmcitethebibliography\endthebibliography}{}
\begin{mcitethebibliography}{83}
\providecommand*\natexlab[1]{#1}
\providecommand*\mciteSetBstSublistMode[1]{}
\providecommand*\mciteSetBstMaxWidthForm[2]{}
\providecommand*\mciteBstWouldAddEndPuncttrue
  {\def\EndOfBibitem{\unskip.}}
\providecommand*\mciteBstWouldAddEndPunctfalse
  {\let\EndOfBibitem\relax}
\providecommand*\mciteSetBstMidEndSepPunct[3]{}
\providecommand*\mciteSetBstSublistLabelBeginEnd[3]{}
\providecommand*\EndOfBibitem{}
\mciteSetBstSublistMode{f}
\mciteSetBstMaxWidthForm{subitem}{(\alph{mcitesubitemcount})}
\mciteSetBstSublistLabelBeginEnd
  {\mcitemaxwidthsubitemform\space}
  {\relax}
  {\relax}

\bibitem[Ghimire and Mazin(2020)Ghimire, and Mazin]{ghimire2020topology}
Ghimire,~N.~J.; Mazin,~I.~I. {Topology and Correlations on the Kagome Lattice}.
  \emph{Nat. Mater.} \textbf{2020}, \emph{19}, 137--138\relax
\mciteBstWouldAddEndPuncttrue
\mciteSetBstMidEndSepPunct{\mcitedefaultmidpunct}
{\mcitedefaultendpunct}{\mcitedefaultseppunct}\relax
\EndOfBibitem
\bibitem[Kang \latin{et~al.}(2020)Kang, Ye, Fang, You, Levitan, Han, Facio,
  Jozwiak, Bostwick, Rotenberg, Chan, McDonald, Graf, Kaznatcheev, Vescovo,
  Bell, Kaxiras, Brink, Richter, Ghimire, \latin{et~al.} others]{kang2020dirac}
Kang,~M.; Ye,~L.; Fang,~S.; You,~J.-S.; Levitan,~A.; Han,~M.; Facio,~J.~I.;
  Jozwiak,~C.; Bostwick,~A.; Rotenberg,~E.; Chan,~M.~K.; McDonald,~R.~D.;
  Graf,~D.; Kaznatcheev,~K.; Vescovo,~E.; Bell,~D.~C.; Kaxiras,~E.; Brink,~J.
  v.~d.; Richter,~M.; Ghimire,~M.~P., \latin{et~al.}  {Dirac Fermions and Flat
  Bands in the Ideal Kagome Metal {{FeSn}}}. \emph{Nat. Mater.} \textbf{2020},
  \emph{19}, 163--169\relax
\mciteBstWouldAddEndPuncttrue
\mciteSetBstMidEndSepPunct{\mcitedefaultmidpunct}
{\mcitedefaultendpunct}{\mcitedefaultseppunct}\relax
\EndOfBibitem
\bibitem[Neupert \latin{et~al.}(2022)Neupert, Denner, Yin, Thomale, and
  Hasan]{neupert2022charge}
Neupert,~T.; Denner,~M.~M.; Yin,~J.-X.; Thomale,~R.; Hasan,~M.~Z. {Charge Order
  and Superconductivity in Kagome Materials}. \emph{Nat. Phys.} \textbf{2022},
  \emph{18}, 137--143\relax
\mciteBstWouldAddEndPuncttrue
\mciteSetBstMidEndSepPunct{\mcitedefaultmidpunct}
{\mcitedefaultendpunct}{\mcitedefaultseppunct}\relax
\EndOfBibitem
\bibitem[Yin \latin{et~al.}(2022)Yin, Lian, and Hasan]{yin2022topological}
Yin,~J.-X.; Lian,~B.; Hasan,~M.~Z. {Topological Kagome Magnets and
  Superconductors}. \emph{Nature} \textbf{2022}, \emph{612}, 647--657\relax
\mciteBstWouldAddEndPuncttrue
\mciteSetBstMidEndSepPunct{\mcitedefaultmidpunct}
{\mcitedefaultendpunct}{\mcitedefaultseppunct}\relax
\EndOfBibitem
\bibitem[Wang \latin{et~al.}(2023)Wang, Wu, McCandless, Chan, and
  Ali]{wang2023quantum}
Wang,~Y.; Wu,~H.; McCandless,~G.~T.; Chan,~J.~Y.; Ali,~M.~N. {Quantum States
  and Intertwining Phases in Kagome Materials}. \emph{Nat. Rev. Phys.}
  \textbf{2023}, \emph{5}, 635--658\relax
\mciteBstWouldAddEndPuncttrue
\mciteSetBstMidEndSepPunct{\mcitedefaultmidpunct}
{\mcitedefaultendpunct}{\mcitedefaultseppunct}\relax
\EndOfBibitem
\bibitem[Kraut and von Baltz(1979)Kraut, and von Baltz]{kraut1979anomalous}
Kraut,~W.; von Baltz,~R. {Anomalous Bulk Photovoltaic Effect in Ferroelectrics:
  A Quadratic Response Theory}. \emph{Phys. Rev. B} \textbf{1979}, \emph{19},
  1548\relax
\mciteBstWouldAddEndPuncttrue
\mciteSetBstMidEndSepPunct{\mcitedefaultmidpunct}
{\mcitedefaultendpunct}{\mcitedefaultseppunct}\relax
\EndOfBibitem
\bibitem[von Baltz and Kraut(1981)von Baltz, and Kraut]{von1981theory}
von Baltz,~R.; Kraut,~W. {Theory of the Bulk Photovoltaic Effect in Pure
  Crystals}. \emph{Phys. Rev. B} \textbf{1981}, \emph{23}, 5590\relax
\mciteBstWouldAddEndPuncttrue
\mciteSetBstMidEndSepPunct{\mcitedefaultmidpunct}
{\mcitedefaultendpunct}{\mcitedefaultseppunct}\relax
\EndOfBibitem
\bibitem[Campi \latin{et~al.}(1998)Campi, Col{\`\i}, and
  Vallone]{campi1998formulation}
Campi,~D.; Col{\`\i},~G.; Vallone,~M. {Formulation of the Optical Response in
  Semiconductors and Quantum-Confined Structures}. \emph{Phys. Rev. B}
  \textbf{1998}, \emph{57}, 4681\relax
\mciteBstWouldAddEndPuncttrue
\mciteSetBstMidEndSepPunct{\mcitedefaultmidpunct}
{\mcitedefaultendpunct}{\mcitedefaultseppunct}\relax
\EndOfBibitem
\bibitem[Orapunt and O’Leary(2008)Orapunt, and O’Leary]{orapunt2008optical}
Orapunt,~F.; O’Leary,~S.~K. {Optical Transitions and the Mobility Edge in
  Amorphous Semiconductors: A Joint Density of States Analysis}. \emph{J. Appl.
  Phys.} \textbf{2008}, \emph{104}, 073513\relax
\mciteBstWouldAddEndPuncttrue
\mciteSetBstMidEndSepPunct{\mcitedefaultmidpunct}
{\mcitedefaultendpunct}{\mcitedefaultseppunct}\relax
\EndOfBibitem
\bibitem[Xiao \latin{et~al.}(2010)Xiao, Chang, and Niu]{xiao2010berry}
Xiao,~D.; Chang,~M.-C.; Niu,~Q. {Berry Phase Effects on Electronic Properties}.
  \emph{Rev. Mod. Phys.} \textbf{2010}, \emph{82}, 1959--2007\relax
\mciteBstWouldAddEndPuncttrue
\mciteSetBstMidEndSepPunct{\mcitedefaultmidpunct}
{\mcitedefaultendpunct}{\mcitedefaultseppunct}\relax
\EndOfBibitem
\bibitem[Young and Rappe(2012)Young, and Rappe]{young2012first}
Young,~S.~M.; Rappe,~A.~M. {First Principles Calculation of the Shift Current
  Photovoltaic Effect in Ferroelectrics}. \emph{Phys. Rev. Lett.}
  \textbf{2012}, \emph{109}, 116601\relax
\mciteBstWouldAddEndPuncttrue
\mciteSetBstMidEndSepPunct{\mcitedefaultmidpunct}
{\mcitedefaultendpunct}{\mcitedefaultseppunct}\relax
\EndOfBibitem
\bibitem[Tan and Rappe(2016)Tan, and Rappe]{tan2016enhancement}
Tan,~L.~Z.; Rappe,~A.~M. {Enhancement of the Bulk Photovoltaic Effect in
  Topological Insulators}. \emph{Phys. Rev. Lett.} \textbf{2016}, \emph{116},
  237402\relax
\mciteBstWouldAddEndPuncttrue
\mciteSetBstMidEndSepPunct{\mcitedefaultmidpunct}
{\mcitedefaultendpunct}{\mcitedefaultseppunct}\relax
\EndOfBibitem
\bibitem[Gupta \latin{et~al.}(2018)Gupta, Shirodkar, Kutana, and
  Yakobson]{gupta2018pursuit}
Gupta,~S.; Shirodkar,~S.~N.; Kutana,~A.; Yakobson,~B.~I. {In Pursuit of {{2D}}
  Materials for Maximum Optical Response}. \emph{ACS Nano} \textbf{2018},
  \emph{12}, 10880--10889\relax
\mciteBstWouldAddEndPuncttrue
\mciteSetBstMidEndSepPunct{\mcitedefaultmidpunct}
{\mcitedefaultendpunct}{\mcitedefaultseppunct}\relax
\EndOfBibitem
\bibitem[Qian \latin{et~al.}(2023)Qian, Zhou, Wang, and Liu]{qian2023shift}
Qian,~Z.; Zhou,~J.; Wang,~H.; Liu,~S. {Shift Current Response in Elemental
  Two-Dimensional Ferroelectrics}. \emph{npj Comput. Mater.} \textbf{2023},
  \emph{9}, 67\relax
\mciteBstWouldAddEndPuncttrue
\mciteSetBstMidEndSepPunct{\mcitedefaultmidpunct}
{\mcitedefaultendpunct}{\mcitedefaultseppunct}\relax
\EndOfBibitem
\bibitem[You \latin{et~al.}(2024)You, Su, and Feng]{you2024versatile}
You,~J.-Y.; Su,~G.; Feng,~Y.~P. {A Versatile Model with Three-Dimensional
  Triangular Lattice for Unconventional Transport and Various Topological
  Effects}. \emph{Natl. Sci. Rev.} \textbf{2024}, \emph{11}, nwad114\relax
\mciteBstWouldAddEndPuncttrue
\mciteSetBstMidEndSepPunct{\mcitedefaultmidpunct}
{\mcitedefaultendpunct}{\mcitedefaultseppunct}\relax
\EndOfBibitem
\bibitem[Tanaka \latin{et~al.}(2024)Tanaka, Watanabe, and
  Yanase]{tanaka2024nonlinear}
Tanaka,~H.; Watanabe,~H.; Yanase,~Y. {Nonlinear Optical Response in
  Superconductors in Magnetic Field: Quantum Geometry and Topological
  Superconductivity}. \emph{Phys. Rev. B} \textbf{2024}, \emph{110},
  014520\relax
\mciteBstWouldAddEndPuncttrue
\mciteSetBstMidEndSepPunct{\mcitedefaultmidpunct}
{\mcitedefaultendpunct}{\mcitedefaultseppunct}\relax
\EndOfBibitem
\bibitem[Lai \latin{et~al.}()Lai, Zhan, Liu, Shirakawa, Seiji, Chen, and
  Sun]{laiuniversal}
Lai,~J.; Zhan,~J.; Liu,~P.; Shirakawa,~T.; Seiji,~Y.; Chen,~X.-Q.; Sun,~Y.
  {Universal Enhancement Effect of Nonlinear Optical Response From Band
  Hybridization}. \emph{Adv. Opt. Mater.} 2401143\relax
\mciteBstWouldAddEndPuncttrue
\mciteSetBstMidEndSepPunct{\mcitedefaultmidpunct}
{\mcitedefaultendpunct}{\mcitedefaultseppunct}\relax
\EndOfBibitem
\bibitem[Chen \latin{et~al.}(2024)Chen, Fang, Cheng, L{\"u}, Cao, Zhu, and
  Wu]{chen2024large}
Chen,~Z.; Fang,~Y.; Cheng,~M.; L{\"u},~T.-Y.; Cao,~X.; Zhu,~Z.-Z.; Wu,~S.
  {Large Second-Harmonic Generation and Linear Electro-Optic Effect in the Bulk
  Kagome Lattice Compound Nb$_3$MX$_7$ (M= Se, S, Te; X= I, Br)}. \emph{Phy.
  Rev. B} \textbf{2024}, \emph{109}, 115118\relax
\mciteBstWouldAddEndPuncttrue
\mciteSetBstMidEndSepPunct{\mcitedefaultmidpunct}
{\mcitedefaultendpunct}{\mcitedefaultseppunct}\relax
\EndOfBibitem
\bibitem[Zhang \latin{et~al.}(2020)Zhang, Xia, Wang, Jin, Tian, kin Ho, Xu,
  Jin, and Xie]{zhang2020single}
Zhang,~J.; Xia,~Y.; Wang,~B.; Jin,~Y.; Tian,~H.; kin Ho,~W.; Xu,~H.; Jin,~C.;
  Xie,~M. {Single-Layer {{Mo$_5$Te$_8$}}―A New Polymorph of Layered
  Transition-Metal Chalcogenide}. \emph{2D Mater.} \textbf{2020}, \emph{8},
  015006\relax
\mciteBstWouldAddEndPuncttrue
\mciteSetBstMidEndSepPunct{\mcitedefaultmidpunct}
{\mcitedefaultendpunct}{\mcitedefaultseppunct}\relax
\EndOfBibitem
\bibitem[Wang and Bobev(2023)Wang, and Bobev]{wang2023synthesis}
Wang,~Y.; Bobev,~S. {Synthesis and Crystal Structure of the Zintl Phases
  NaSrSb, NaBaSb and NaEuSb}. \emph{Materials} \textbf{2023}, \emph{16},
  1428\relax
\mciteBstWouldAddEndPuncttrue
\mciteSetBstMidEndSepPunct{\mcitedefaultmidpunct}
{\mcitedefaultendpunct}{\mcitedefaultseppunct}\relax
\EndOfBibitem
\bibitem[Iandelli and Franceschi(1973)Iandelli, and
  Franceschi]{iandelli1973crystal}
Iandelli,~A.; Franceschi,~E. {On the Crystal Structure of the Compounds CaP,
  SrP, CaAs, SrAs and EuAs}. \emph{J. Less Common Met.} \textbf{1973},
  \emph{30}, 211--216\relax
\mciteBstWouldAddEndPuncttrue
\mciteSetBstMidEndSepPunct{\mcitedefaultmidpunct}
{\mcitedefaultendpunct}{\mcitedefaultseppunct}\relax
\EndOfBibitem
\bibitem[Zhuravlev and Obolonskaya(2010)Zhuravlev, and
  Obolonskaya]{zhuravlev2010structure}
Zhuravlev,~Y.~N.; Obolonskaya,~O. {Structure, Mechanical Stability, and
  Chemical Bond in Alkali Metal Oxides}. \emph{J. Struct. Chem.} \textbf{2010},
  \emph{51}, 1005--1013\relax
\mciteBstWouldAddEndPuncttrue
\mciteSetBstMidEndSepPunct{\mcitedefaultmidpunct}
{\mcitedefaultendpunct}{\mcitedefaultseppunct}\relax
\EndOfBibitem
\bibitem[Duan \latin{et~al.}(2024)Duan, You, Cai, Gou, Li, Huang, Yu, Teo, Sun,
  Wang, \latin{et~al.} others]{duan2024observation}
Duan,~S.; You,~J.-Y.; Cai,~Z.; Gou,~J.; Li,~D.; Huang,~Y.~L.; Yu,~X.;
  Teo,~S.~L.; Sun,~S.; Wang,~Y., \latin{et~al.}  {Observation of Kagome-Like
  Bands in Two-Dimensional Semiconducting {{Cr$_8$Se$_{{12}}$}}}. \emph{Nat.
  Commun.} \textbf{2024}, \emph{15}, 8940\relax
\mciteBstWouldAddEndPuncttrue
\mciteSetBstMidEndSepPunct{\mcitedefaultmidpunct}
{\mcitedefaultendpunct}{\mcitedefaultseppunct}\relax
\EndOfBibitem
\bibitem[Xie \latin{et~al.}(2024)Xie, Ji, He, Shen, Wang, and
  Zhang]{xie2024manipulation}
Xie,~Y.; Ji,~K.; He,~J.; Shen,~X.; Wang,~D.; Zhang,~J. {Manipulation of
  Topology by Electric Field in Breathing Kagome Lattice}. \emph{arXiv}
  \textbf{2024}, 2411.17208,
  DOI:{\color{blue}{10.48550/arXiv.2411.17208}}\relax
\mciteBstWouldAddEndPuncttrue
\mciteSetBstMidEndSepPunct{\mcitedefaultmidpunct}
{\mcitedefaultendpunct}{\mcitedefaultseppunct}\relax
\EndOfBibitem
\bibitem[Du \latin{et~al.}(2021)Du, Hasan, Castellanos-Gomez, Liu, Yao, Lau,
  and Sun]{du2021engineering}
Du,~L.; Hasan,~T.; Castellanos-Gomez,~A.; Liu,~G.-B.; Yao,~Y.; Lau,~C.~N.;
  Sun,~Z. {Engineering Symmetry Breaking in 2D Layered Materials}. \emph{Nat.
  Rev. Phys.} \textbf{2021}, \emph{3}, 193--206\relax
\mciteBstWouldAddEndPuncttrue
\mciteSetBstMidEndSepPunct{\mcitedefaultmidpunct}
{\mcitedefaultendpunct}{\mcitedefaultseppunct}\relax
\EndOfBibitem
\bibitem[Zhou \latin{et~al.}(2024)Zhou, dos Santos~Dias, Zhang, Zhao, and
  Lounis]{zhou2024kagomerization}
Zhou,~H.; dos Santos~Dias,~M.; Zhang,~Y.; Zhao,~W.; Lounis,~S. {Kagomerization
  of Transition Metal Monolayers Induced by Two-Dimensional Hexagonal Boron
  Nitride}. \emph{Nat. Commun.} \textbf{2024}, \emph{15}, 4854\relax
\mciteBstWouldAddEndPuncttrue
\mciteSetBstMidEndSepPunct{\mcitedefaultmidpunct}
{\mcitedefaultendpunct}{\mcitedefaultseppunct}\relax
\EndOfBibitem
\bibitem[Deng \latin{et~al.}(2025)Deng, Guo, Wen, Lu, Zhang, Cheng, Pan, Jian,
  Li, Wang, Bai, Li, Ji, He, and Zhang]{deng2024ferroelectricity}
Deng,~J.; Guo,~D.; Wen,~Y.; Lu,~S.; Zhang,~H.; Cheng,~Z.; Pan,~Z.; Jian,~T.;
  Li,~D.; Wang,~H.; Bai,~Y.; Li,~Z.; Ji,~W.; He,~j.; Zhang,~C. {Evidence of
  Ferroelectricity in an Antiferromagnetic Vanadium Trichloride Monolayer}.
  \emph{Science Advances} \textbf{2025}, \emph{11}, eado6538\relax
\mciteBstWouldAddEndPuncttrue
\mciteSetBstMidEndSepPunct{\mcitedefaultmidpunct}
{\mcitedefaultendpunct}{\mcitedefaultseppunct}\relax
\EndOfBibitem
\bibitem[Lee \latin{et~al.}(2024)Lee, Kim, Song, Kim, Lee, Yoo, Cho, Rhim,
  Jung, Kim, and Kim]{lee2023unconventional}
Lee,~J.~H.; Kim,~G.~W.; Song,~I.; Kim,~Y.; Lee,~Y.; Yoo,~S.~J.; Cho,~D.-Y.;
  Rhim,~J.-W.; Jung,~J.; Kim,~G.; Kim,~C. {Atomically Thin Two-Dimensional
  Kagome Flat Band on the Silicon Surface}. \emph{ACS nano} \textbf{2024},
  \emph{18}, 25535--25541\relax
\mciteBstWouldAddEndPuncttrue
\mciteSetBstMidEndSepPunct{\mcitedefaultmidpunct}
{\mcitedefaultendpunct}{\mcitedefaultseppunct}\relax
\EndOfBibitem
\bibitem[Li \latin{et~al.}(2024)Li, Zhai, Liu, Zhang, Meng, Zhuang, Feng, Xu,
  Hao, Zhou, Lu, Dou, and Du]{li2024electronic}
Li,~Y.; Zhai,~S.; Liu,~Y.; Zhang,~J.; Meng,~Z.; Zhuang,~J.; Feng,~H.; Xu,~X.;
  Hao,~W.; Zhou,~M.; Lu,~G.-H.; Dou,~S.; Du,~Y. {Electronic Flat Band in
  Distorted Colouring Triangle Lattice}. \emph{Adv. Sci.} \textbf{2024},
  \emph{11}, 2303483\relax
\mciteBstWouldAddEndPuncttrue
\mciteSetBstMidEndSepPunct{\mcitedefaultmidpunct}
{\mcitedefaultendpunct}{\mcitedefaultseppunct}\relax
\EndOfBibitem
\bibitem[Liu \latin{et~al.}(2024)Liu, Chang, Wang, Zhou, Wang, Fan, Han, Li,
  Ren, Wang, Chen, and Zhang]{liu2024cascade}
Liu,~C.; Chang,~T.; Wang,~S.; Zhou,~S.; Wang,~X.; Fan,~C.; Han,~L.; Li,~F.;
  Ren,~H.; Wang,~S.; Chen,~Y.-S.; Zhang,~J. {Cascade of Phase Transitions and
  Large Magnetic Anisotropy in a Triangle-Kagome-Triangle Trilayer
  Antiferromagnet}. \emph{Chem. Mater.} \textbf{2024}, \emph{36},
  9516--9525\relax
\mciteBstWouldAddEndPuncttrue
\mciteSetBstMidEndSepPunct{\mcitedefaultmidpunct}
{\mcitedefaultendpunct}{\mcitedefaultseppunct}\relax
\EndOfBibitem
\bibitem[Wang \latin{et~al.}(2022)Wang, McCandless, Wang, Thanabalasingam, Wu,
  Bouwmeester, Van Der~Zant, Ali, and Chan]{wang2022electronic}
Wang,~Y.; McCandless,~G.~T.; Wang,~X.; Thanabalasingam,~K.; Wu,~H.;
  Bouwmeester,~D.; Van Der~Zant,~H.~S.; Ali,~M.~N.; Chan,~J.~Y. {Electronic
  Properties and Phase Transition in the Kagome Metal Yb$_{{0.
  5}}$Co$_{{3}}$Ge$_{{3}}$}. \emph{Chem. Mater.} \textbf{2022}, \emph{34},
  7337--7343\relax
\mciteBstWouldAddEndPuncttrue
\mciteSetBstMidEndSepPunct{\mcitedefaultmidpunct}
{\mcitedefaultendpunct}{\mcitedefaultseppunct}\relax
\EndOfBibitem
\bibitem[Wu \latin{et~al.}(2024)Wu, Klemm, Shah, Ritz, Duan, Teng, Gao, Ye,
  Matsuda, Li, Xu, Yi, Birol, Dai, and Blumberg]{wu2024symmetry}
Wu,~S.; Klemm,~M.~L.; Shah,~J.; Ritz,~E.~T.; Duan,~C.; Teng,~X.; Gao,~B.;
  Ye,~F.; Matsuda,~M.; Li,~F.; Xu,~X.; Yi,~M.; Birol,~T.; Dai,~P.; Blumberg,~G.
  {Symmetry Breaking and Ascending in the Magnetic Kagome Metal FeGe}.
  \emph{Phys. Rev. X} \textbf{2024}, \emph{14}, 011043\relax
\mciteBstWouldAddEndPuncttrue
\mciteSetBstMidEndSepPunct{\mcitedefaultmidpunct}
{\mcitedefaultendpunct}{\mcitedefaultseppunct}\relax
\EndOfBibitem
\bibitem[Li \latin{et~al.}(2019)Li, Yu, Shen, Tang, and Han]{li2019external}
Li,~H.; Yu,~X.; Shen,~X.; Tang,~G.; Han,~K. {External Electric Field Induced
  Second-Order Nonlinear Optical Effects in Hexagonal Graphene Quantum Dots}.
  \emph{J. Phys. Chem. C} \textbf{2019}, \emph{123}, 20020--20025\relax
\mciteBstWouldAddEndPuncttrue
\mciteSetBstMidEndSepPunct{\mcitedefaultmidpunct}
{\mcitedefaultendpunct}{\mcitedefaultseppunct}\relax
\EndOfBibitem
\bibitem[Jiang \latin{et~al.}(2021)Jiang, Chen, Hu, Xiang, Zhang, Wang, Wang,
  and Shi]{jiang2021flexo}
Jiang,~J.; Chen,~Z.; Hu,~Y.; Xiang,~Y.; Zhang,~L.; Wang,~Y.; Wang,~G.-C.;
  Shi,~J. {Flexo-Photovoltaic Effect in MoS$_2$}. \emph{Nat. Nanotechnol.}
  \textbf{2021}, \emph{16}, 894--901\relax
\mciteBstWouldAddEndPuncttrue
\mciteSetBstMidEndSepPunct{\mcitedefaultmidpunct}
{\mcitedefaultendpunct}{\mcitedefaultseppunct}\relax
\EndOfBibitem
\bibitem[Zhang \latin{et~al.}(2023)Zhang, Oli, Zou, Guo, Wang, and
  Li]{zhang2023visualizing}
Zhang,~H.; Oli,~B.~D.; Zou,~Q.; Guo,~X.; Wang,~Z.; Li,~L. {Visualizing
  Symmetry-Breaking Electronic Orders in Epitaxial Kagome Magnet FeSn Films}.
  \emph{Nat. Commun.} \textbf{2023}, \emph{14}, 6167\relax
\mciteBstWouldAddEndPuncttrue
\mciteSetBstMidEndSepPunct{\mcitedefaultmidpunct}
{\mcitedefaultendpunct}{\mcitedefaultseppunct}\relax
\EndOfBibitem
\bibitem[Zeng \latin{et~al.}(2020)Zeng, Zhang, Sui, Li, OuYang, Pu, Chen, Ma,
  Cheng, Yan, Xu, and Hong]{zeng2020inversion}
Zeng,~G.; Zhang,~R.; Sui,~Y.; Li,~X.; OuYang,~H.; Pu,~M.; Chen,~H.; Ma,~X.;
  Cheng,~X.; Yan,~W.; Xu,~M.; Hong,~M. {Inversion Symmetry Breaking in Lithium
  Intercalated Graphitic Materials}. \emph{ACS Applied Materials \& Interfaces}
  \textbf{2020}, \emph{12}, 28561--28567\relax
\mciteBstWouldAddEndPuncttrue
\mciteSetBstMidEndSepPunct{\mcitedefaultmidpunct}
{\mcitedefaultendpunct}{\mcitedefaultseppunct}\relax
\EndOfBibitem
\bibitem[Wu \latin{et~al.}(2024)Wu, Quan, Pan, Hu, Zhang, Wang, Zheng, and
  Zhang]{wu2024atomically}
Wu,~Q.; Quan,~W.; Pan,~S.; Hu,~J.; Zhang,~Z.; Wang,~J.; Zheng,~F.; Zhang,~Y.
  {Atomically Thin Kagome-Structured {{Co$_9$Te$_{{16}}$}} Achieved Through
  Self-Intercalation and Its Flat Band Visualization}. \emph{Nano Lett.}
  \textbf{2024}, \emph{24}, 7672--7680\relax
\mciteBstWouldAddEndPuncttrue
\mciteSetBstMidEndSepPunct{\mcitedefaultmidpunct}
{\mcitedefaultendpunct}{\mcitedefaultseppunct}\relax
\EndOfBibitem
\bibitem[Jo \latin{et~al.}(2022)Jo, Kim, Kim, Lee, Choe, Oh, Lee, Lee, Jin, and
  Yoo]{jo2022defect}
Jo,~J.; Kim,~J.~H.; Kim,~C.~H.; Lee,~J.; Choe,~D.; Oh,~I.; Lee,~S.; Lee,~Z.;
  Jin,~H.; Yoo,~J.-W. {Defect-Gradient-Induced Rashba Effect in Van Der Waals
  {{PtSe$_2$}} Layers}. \emph{Nat. Commun.} \textbf{2022}, \emph{13},
  2759\relax
\mciteBstWouldAddEndPuncttrue
\mciteSetBstMidEndSepPunct{\mcitedefaultmidpunct}
{\mcitedefaultendpunct}{\mcitedefaultseppunct}\relax
\EndOfBibitem
\bibitem[Mu \latin{et~al.}(2023)Mu, Xue, Sun, and Zhou]{mu2023magnetic}
Mu,~X.; Xue,~Q.; Sun,~Y.; Zhou,~J. {Magnetic Proximity Enabled Bulk
  Photovoltaic Effects in Van Der Waals Heterostructures}. \emph{Phys. Rev.
  Res.} \textbf{2023}, \emph{5}, 013001\relax
\mciteBstWouldAddEndPuncttrue
\mciteSetBstMidEndSepPunct{\mcitedefaultmidpunct}
{\mcitedefaultendpunct}{\mcitedefaultseppunct}\relax
\EndOfBibitem
\bibitem[Gao \latin{et~al.}(2024)Gao, Yan, Hu, and Chen]{gao2024bilayer}
Gao,~Q.; Yan,~Q.; Hu,~Z.; Chen,~L. {Bilayer Kagome Borophene with Multiple Van
  Hove Singularities}. \emph{Adv. Sci.} \textbf{2024}, \emph{11}, 2305059\relax
\mciteBstWouldAddEndPuncttrue
\mciteSetBstMidEndSepPunct{\mcitedefaultmidpunct}
{\mcitedefaultendpunct}{\mcitedefaultseppunct}\relax
\EndOfBibitem
\bibitem[Ziman(1971)]{ziman1971calculation}
Ziman,~J. \emph{Solid State Physics}; Elsevier, 1971; Vol.~26; pp 1--101\relax
\mciteBstWouldAddEndPuncttrue
\mciteSetBstMidEndSepPunct{\mcitedefaultmidpunct}
{\mcitedefaultendpunct}{\mcitedefaultseppunct}\relax
\EndOfBibitem
\bibitem[Anderson \latin{et~al.}(1974)Anderson, Clunie, and
  Pommerenke]{pommerenke1974bloch}
Anderson,~J.~M.; Clunie,~J.; Pommerenke,~C. {On Bloch Functions and Normal
  Functions.} \emph{J. Reine Angew. Math.} \textbf{1974}, \emph{270},
  12--37\relax
\mciteBstWouldAddEndPuncttrue
\mciteSetBstMidEndSepPunct{\mcitedefaultmidpunct}
{\mcitedefaultendpunct}{\mcitedefaultseppunct}\relax
\EndOfBibitem
\bibitem[Beugeling \latin{et~al.}(2012)Beugeling, Everts, and
  Morais~Smith]{beugeling2012topological}
Beugeling,~W.; Everts,~J.; Morais~Smith,~C. {Topological Phase Transitions
  Driven by Next-Nearest-Neighbor Hopping in Two-Dimensional Lattices}.
  \emph{Phys. Rev. B} \textbf{2012}, \emph{86}, 195129\relax
\mciteBstWouldAddEndPuncttrue
\mciteSetBstMidEndSepPunct{\mcitedefaultmidpunct}
{\mcitedefaultendpunct}{\mcitedefaultseppunct}\relax
\EndOfBibitem
\bibitem[Cook \latin{et~al.}(2017)Cook, M.~Fregoso, De~Juan, Coh, and
  Moore]{cook2017design}
Cook,~A.~M.; M.~Fregoso,~B.; De~Juan,~F.; Coh,~S.; Moore,~J.~E. {Design
  Principles for Shift Current Photovoltaics}. \emph{Nat. Commun.}
  \textbf{2017}, \emph{8}, 14176\relax
\mciteBstWouldAddEndPuncttrue
\mciteSetBstMidEndSepPunct{\mcitedefaultmidpunct}
{\mcitedefaultendpunct}{\mcitedefaultseppunct}\relax
\EndOfBibitem
\bibitem[Zhang \latin{et~al.}(2019)Zhang, Kang, Huang, Jiang, Ni, Kang, Zhang,
  Xu, Liu, and Liu]{zhang2019kagome}
Zhang,~S.; Kang,~M.; Huang,~H.; Jiang,~W.; Ni,~X.; Kang,~L.; Zhang,~S.; Xu,~H.;
  Liu,~Z.; Liu,~F. {Kagome Bands Disguised in a Coloring-Triangle Lattice}.
  \emph{Phys. Rev. B} \textbf{2019}, \emph{99}, 100404\relax
\mciteBstWouldAddEndPuncttrue
\mciteSetBstMidEndSepPunct{\mcitedefaultmidpunct}
{\mcitedefaultendpunct}{\mcitedefaultseppunct}\relax
\EndOfBibitem
\bibitem[Iba{\~n}ez-Azpiroz \latin{et~al.}(2018)Iba{\~n}ez-Azpiroz, Tsirkin,
  and Souza]{ibanez2018ab}
Iba{\~n}ez-Azpiroz,~J.; Tsirkin,~S.~S.; Souza,~I. {Ab Initio Calculation of the
  Shift Photocurrent by Wannier Interpolation}. \emph{Phys. Rev. B}
  \textbf{2018}, \emph{97}, 245143\relax
\mciteBstWouldAddEndPuncttrue
\mciteSetBstMidEndSepPunct{\mcitedefaultmidpunct}
{\mcitedefaultendpunct}{\mcitedefaultseppunct}\relax
\EndOfBibitem
\bibitem[Zhang \latin{et~al.}(2019)Zhang, Ideue, Onga, Qin, Suzuki, Zak, Tenne,
  Smet, and Iwasa]{zhang2019enhanced}
Zhang,~Y.; Ideue,~T.; Onga,~M.; Qin,~F.; Suzuki,~R.; Zak,~A.; Tenne,~R.;
  Smet,~J.; Iwasa,~Y. {Enhanced Intrinsic Photovoltaic Effect in Tungsten
  Disulfide Nanotubes}. \emph{Nature} \textbf{2019}, \emph{570}, 349--353\relax
\mciteBstWouldAddEndPuncttrue
\mciteSetBstMidEndSepPunct{\mcitedefaultmidpunct}
{\mcitedefaultendpunct}{\mcitedefaultseppunct}\relax
\EndOfBibitem
\bibitem[Sauer \latin{et~al.}(2023)Sauer, Taghizadeh, Petralanda, Ovesen,
  Thygesen, Olsen, Cornean, and Pedersen]{sauer2023shift2}
Sauer,~M.~O.; Taghizadeh,~A.; Petralanda,~U.; Ovesen,~M.; Thygesen,~K.~S.;
  Olsen,~T.; Cornean,~H.; Pedersen,~T.~G. {Shift Current Photovoltaic
  Efficiency of {{2D}} Materials}. \emph{npj Comput. Mater.} \textbf{2023},
  \emph{9}, 35\relax
\mciteBstWouldAddEndPuncttrue
\mciteSetBstMidEndSepPunct{\mcitedefaultmidpunct}
{\mcitedefaultendpunct}{\mcitedefaultseppunct}\relax
\EndOfBibitem
\bibitem[Liang \latin{et~al.}(2023)Liang, Zhou, Zhang, Yu, Lv, Song, Zhou,
  Wang, Wang, Wang, Shum, He, Liu, Zhu, Wang, and Chen]{liang2023strong}
Liang,~Z.; Zhou,~X.; Zhang,~L.; Yu,~X.-L.; Lv,~Y.; Song,~X.; Zhou,~Y.;
  Wang,~H.; Wang,~S.; Wang,~T.; Shum,~P.~P.; He,~Q.; Liu,~Y.; Zhu,~C.;
  Wang,~L.; Chen,~X. {Strong Bulk Photovoltaic Effect in Engineered
  Edge-Embedded Van Der Waals Structures}. \emph{Nat. Commun.} \textbf{2023},
  \emph{14}, 4230\relax
\mciteBstWouldAddEndPuncttrue
\mciteSetBstMidEndSepPunct{\mcitedefaultmidpunct}
{\mcitedefaultendpunct}{\mcitedefaultseppunct}\relax
\EndOfBibitem
\bibitem[Lei \latin{et~al.}(2023)Lei, Dai, Dong, Geng, Cao, Wang, Xu, Pang,
  Liu, Li, Cheng, Wang, and Ji]{lei2023electronic}
Lei,~L.; Dai,~J.; Dong,~H.; Geng,~Y.; Cao,~F.; Wang,~C.; Xu,~R.; Pang,~F.;
  Liu,~Z.-X.; Li,~F.; Cheng,~Z.; Wang,~G.; Ji,~W. {Electronic Janus Lattice and
  Kagome-Like Bands in Coloring-Triangular {{MoTe$_2$}} Monolayers}. \emph{Nat.
  Commun.} \textbf{2023}, \emph{14}, 6320\relax
\mciteBstWouldAddEndPuncttrue
\mciteSetBstMidEndSepPunct{\mcitedefaultmidpunct}
{\mcitedefaultendpunct}{\mcitedefaultseppunct}\relax
\EndOfBibitem
\bibitem[Tan \latin{et~al.}(2016)Tan, Zheng, Young, Wang, Liu, and
  Rappe]{tan2016shift}
Tan,~L.~Z.; Zheng,~F.; Young,~S.~M.; Wang,~F.; Liu,~S.; Rappe,~A.~M. {Shift
  Current Bulk Photovoltaic Effect in Polar Materials—Hybrid and Oxide
  Perovskites and Beyond}. \emph{npj Comput. Mater.} \textbf{2016}, \emph{2},
  16026\relax
\mciteBstWouldAddEndPuncttrue
\mciteSetBstMidEndSepPunct{\mcitedefaultmidpunct}
{\mcitedefaultendpunct}{\mcitedefaultseppunct}\relax
\EndOfBibitem
\bibitem[Bihari~Swain \latin{et~al.}(2019)Bihari~Swain, Murali, Nanda, and
  Murugavel]{bihari2019large}
Bihari~Swain,~A.; Murali,~D.; Nanda,~B.; Murugavel,~P. {Large Bulk Photovoltaic
  Response by Symmetry-Breaking Structural Transformation in Ferroelectric
  {{[Ba(Zr$_{{0.2}}$Ti$_{{0.8}}$)O$_3$]$_{{0.5}}$[(Ba$_{{0.7}}$Ca$_{{0.3}}$)TiO$_3$]$_{{0.5}}$}}}.
  \emph{Phys. Rev. Appl.} \textbf{2019}, \emph{11}, 044007\relax
\mciteBstWouldAddEndPuncttrue
\mciteSetBstMidEndSepPunct{\mcitedefaultmidpunct}
{\mcitedefaultendpunct}{\mcitedefaultseppunct}\relax
\EndOfBibitem
\bibitem[Tan and Rappe(2019)Tan, and Rappe]{tan2019upper}
Tan,~L.~Z.; Rappe,~A.~M. {Upper Limit on Shift Current Generation in Extended
  Systems}. \emph{Phys. Rev. B} \textbf{2019}, \emph{100}, 085102\relax
\mciteBstWouldAddEndPuncttrue
\mciteSetBstMidEndSepPunct{\mcitedefaultmidpunct}
{\mcitedefaultendpunct}{\mcitedefaultseppunct}\relax
\EndOfBibitem
\bibitem[Dai \latin{et~al.}(2024)Dai, Zhang, Pan, Wang, Zhang, Cheng, and
  Ji]{dai2024kagome}
Dai,~J.; Zhang,~Z.; Pan,~Z.; Wang,~C.; Zhang,~C.; Cheng,~Z.; Ji,~W. {Kagome
  Bands and Magnetism in MoTe$_{2-x}$ Kagome Monolayers}. \emph{arXiv}
  \textbf{2024}, 2408.14285,
  DOI:{\color{blue}{10.48550/arXiv.2408.14285}}\relax
\mciteBstWouldAddEndPuncttrue
\mciteSetBstMidEndSepPunct{\mcitedefaultmidpunct}
{\mcitedefaultendpunct}{\mcitedefaultseppunct}\relax
\EndOfBibitem
\bibitem[Hong \latin{et~al.}(2017)Hong, Wang, Liu, Ren, Chen, Wang, Jia, Xie,
  Jin, Ji, Yuan, and Zhang]{hong2017inversion}
Hong,~J.; Wang,~C.; Liu,~H.; Ren,~X.; Chen,~J.; Wang,~G.; Jia,~J.; Xie,~M.;
  Jin,~C.; Ji,~W.; Yuan,~J.; Zhang,~Z. {Inversion Domain Boundary Induced
  Stacking and Bandstructure Diversity in Bilayer {{MoSe$_2$}}}. \emph{Nano
  Lett.} \textbf{2017}, \emph{17}, 6653--6660\relax
\mciteBstWouldAddEndPuncttrue
\mciteSetBstMidEndSepPunct{\mcitedefaultmidpunct}
{\mcitedefaultendpunct}{\mcitedefaultseppunct}\relax
\EndOfBibitem
\bibitem[Liu \latin{et~al.}(2014)Liu, Jiao, Yang, Cai, Wu, Ho, Gao, Jia, Wang,
  Fan, Yao, and Xie]{liu2014dense}
Liu,~H.; Jiao,~L.; Yang,~F.; Cai,~Y.; Wu,~X.; Ho,~W.; Gao,~C.; Jia,~J.;
  Wang,~N.; Fan,~H.; Yao,~W.; Xie,~M. {Dense Network of One-Dimensional Midgap
  Metallic Modes in Monolayer {{MoSe$_2$}} and Their Spatial Undulations}.
  \emph{Phys. Rev. Lett.} \textbf{2014}, \emph{113}, 066105\relax
\mciteBstWouldAddEndPuncttrue
\mciteSetBstMidEndSepPunct{\mcitedefaultmidpunct}
{\mcitedefaultendpunct}{\mcitedefaultseppunct}\relax
\EndOfBibitem
\bibitem[Rangel \latin{et~al.}(2017)Rangel, Fregoso, Mendoza, Morimoto, Moore,
  and Neaton]{rangel2017large}
Rangel,~T.; Fregoso,~B.~M.; Mendoza,~B.~S.; Morimoto,~T.; Moore,~J.~E.;
  Neaton,~J.~B. {Large Bulk Photovoltaic Effect and Spontaneous Polarization of
  Single-Layer Monochalcogenides}. \emph{Phys. Rev. Lett.} \textbf{2017},
  \emph{119}, 067402\relax
\mciteBstWouldAddEndPuncttrue
\mciteSetBstMidEndSepPunct{\mcitedefaultmidpunct}
{\mcitedefaultendpunct}{\mcitedefaultseppunct}\relax
\EndOfBibitem
\bibitem[Chan \latin{et~al.}(2021)Chan, Qiu, da~Jornada, and
  Louie]{chan2021giant}
Chan,~Y.-H.; Qiu,~D.~Y.; da~Jornada,~F.~H.; Louie,~S.~G. {Giant
  Exciton-Enhanced Shift Currents and Direct Current Conduction with Subbandgap
  Photo Excitations Produced by Many-Electron Interactions}. \emph{Proc. Natl.
  Acad. Sci.} \textbf{2021}, \emph{118}, e1906938118\relax
\mciteBstWouldAddEndPuncttrue
\mciteSetBstMidEndSepPunct{\mcitedefaultmidpunct}
{\mcitedefaultendpunct}{\mcitedefaultseppunct}\relax
\EndOfBibitem
\bibitem[Xu \latin{et~al.}(2021)Xu, Wang, Zhou, Guo, Kong, and
  Li]{xu2021colossal}
Xu,~H.; Wang,~H.; Zhou,~J.; Guo,~Y.; Kong,~J.; Li,~J. {Colossal Switchable
  Photocurrents in Topological Janus Transition Metal Dichalcogenides}.
  \emph{npj Comput. Mater.} \textbf{2021}, \emph{7}, 31\relax
\mciteBstWouldAddEndPuncttrue
\mciteSetBstMidEndSepPunct{\mcitedefaultmidpunct}
{\mcitedefaultendpunct}{\mcitedefaultseppunct}\relax
\EndOfBibitem
\bibitem[Xiao \latin{et~al.}(2022)Xiao, Gao, Jiang, Gan, Zhang, and
  Li]{xiao2022non}
Xiao,~R.-C.; Gao,~Y.; Jiang,~H.; Gan,~W.; Zhang,~C.; Li,~H. {Non-Synchronous
  Bulk Photovoltaic Effect in Two-Dimensional Interlayer-Sliding
  Ferroelectrics}. \emph{npj Comput. Mater.} \textbf{2022}, \emph{8}, 138\relax
\mciteBstWouldAddEndPuncttrue
\mciteSetBstMidEndSepPunct{\mcitedefaultmidpunct}
{\mcitedefaultendpunct}{\mcitedefaultseppunct}\relax
\EndOfBibitem
\bibitem[Jin and He(2024)Jin, and He]{jin2024peculiar}
Jin,~G.; He,~L. {Peculiar Band Geometry Induced Giant Shift Current in
  Ferroelectric {{SnTe}} Monolayer}. \emph{npj Comput. Mater.} \textbf{2024},
  \emph{10}, 23\relax
\mciteBstWouldAddEndPuncttrue
\mciteSetBstMidEndSepPunct{\mcitedefaultmidpunct}
{\mcitedefaultendpunct}{\mcitedefaultseppunct}\relax
\EndOfBibitem
\bibitem[Young \latin{et~al.}(2012)Young, Zheng, and Rappe]{young2012first2}
Young,~S.~M.; Zheng,~F.; Rappe,~A.~M. {First-Principles Calculation of the Bulk
  Photovoltaic Effect in Bismuth Ferrite}. \emph{Phys. Rev. Lett.}
  \textbf{2012}, \emph{109}, 236601\relax
\mciteBstWouldAddEndPuncttrue
\mciteSetBstMidEndSepPunct{\mcitedefaultmidpunct}
{\mcitedefaultendpunct}{\mcitedefaultseppunct}\relax
\EndOfBibitem
\bibitem[Brehm \latin{et~al.}(2014)Brehm, Young, Zheng, and
  Rappe]{brehm2014first}
Brehm,~J.~A.; Young,~S.~M.; Zheng,~F.; Rappe,~A.~M. {First-Principles
  Calculation of the Bulk Photovoltaic Effect in the Polar Compounds LiAsS$_2$,
  LiAsSe$_2$, and NaAsSe$_2$}. \emph{J. Chem. Phys.} \textbf{2014}, \emph{141},
  204704\relax
\mciteBstWouldAddEndPuncttrue
\mciteSetBstMidEndSepPunct{\mcitedefaultmidpunct}
{\mcitedefaultendpunct}{\mcitedefaultseppunct}\relax
\EndOfBibitem
\bibitem[Wang \latin{et~al.}(2017)Wang, Liu, Kang, Gu, Xu, and
  Duan]{wang2017first}
Wang,~C.; Liu,~X.; Kang,~L.; Gu,~B.-L.; Xu,~Y.; Duan,~W. {First-Principles
  Calculation of Nonlinear Optical Responses by Wannier Interpolation}.
  \emph{Phys. Rev. B} \textbf{2017}, \emph{96}, 115147\relax
\mciteBstWouldAddEndPuncttrue
\mciteSetBstMidEndSepPunct{\mcitedefaultmidpunct}
{\mcitedefaultendpunct}{\mcitedefaultseppunct}\relax
\EndOfBibitem
\bibitem[Zhang \latin{et~al.}(2019)Zhang, Holder, Ishizuka, de~Juan, Nagaosa,
  Felser, and Yan]{zhang2019switchable}
Zhang,~Y.; Holder,~T.; Ishizuka,~H.; de~Juan,~F.; Nagaosa,~N.; Felser,~C.;
  Yan,~B. {Switchable Magnetic Bulk Photovoltaic Effect in the Two-Dimensional
  Magnet {{CrI$_3$}}}. \emph{Nat. Commun.} \textbf{2019}, \emph{10}, 3783\relax
\mciteBstWouldAddEndPuncttrue
\mciteSetBstMidEndSepPunct{\mcitedefaultmidpunct}
{\mcitedefaultendpunct}{\mcitedefaultseppunct}\relax
\EndOfBibitem
\bibitem[Sotome \latin{et~al.}(2019)Sotome, Nakamura, Fujioka, Ogino, Kaneko,
  Morimoto, Zhang, Kawasaki, Nagaosa, Tokura, and Ogawa]{sotome2019spectral}
Sotome,~M.; Nakamura,~M.; Fujioka,~J.; Ogino,~M.; Kaneko,~Y.; Morimoto,~T.;
  Zhang,~Y.; Kawasaki,~M.; Nagaosa,~N.; Tokura,~Y.; Ogawa,~N. {Spectral
  Dynamics of Shift Current in Ferroelectric Semiconductor {{SbSi}}}.
  \emph{Proc. Natl. Acad. Sci.} \textbf{2019}, \emph{116}, 1929--1933\relax
\mciteBstWouldAddEndPuncttrue
\mciteSetBstMidEndSepPunct{\mcitedefaultmidpunct}
{\mcitedefaultendpunct}{\mcitedefaultseppunct}\relax
\EndOfBibitem
\bibitem[Chan \latin{et~al.}(2019)Chan, Qiu, da~Jornada, and
  Louie]{chan2019exciton}
Chan,~Y.-H.; Qiu,~D.~Y.; da~Jornada,~F.~H.; Louie,~S.~G. {Exciton Shift
  Currents: DC Conduction with Sub-Bandgap Photo Excitations}. \emph{arXiv}
  \textbf{2019}, 1904.12813,
  DOI:{\color{blue}{10.48550/arXiv.1904.12813}}\relax
\mciteBstWouldAddEndPuncttrue
\mciteSetBstMidEndSepPunct{\mcitedefaultmidpunct}
{\mcitedefaultendpunct}{\mcitedefaultseppunct}\relax
\EndOfBibitem
\bibitem[Fei \latin{et~al.}(2020)Fei, Tan, and Rappe]{fei2020shift}
Fei,~R.; Tan,~L.~Z.; Rappe,~A.~M. {Shift-Current Bulk Photovoltaic Effect
  Influenced by Quasiparticle and Exciton}. \emph{Phys. Rev. B} \textbf{2020},
  \emph{101}, 045104\relax
\mciteBstWouldAddEndPuncttrue
\mciteSetBstMidEndSepPunct{\mcitedefaultmidpunct}
{\mcitedefaultendpunct}{\mcitedefaultseppunct}\relax
\EndOfBibitem
\bibitem[Schankler \latin{et~al.}(2021)Schankler, Gao, and
  Rappe]{schankler2021large}
Schankler,~A.~M.; Gao,~L.; Rappe,~A.~M. {Large Bulk Piezophotovoltaic Effect of
  Monolayer {{2H-MoS$_2$}}}. \emph{J. Phys. Chem. Lett.} \textbf{2021},
  \emph{12}, 1244--1249\relax
\mciteBstWouldAddEndPuncttrue
\mciteSetBstMidEndSepPunct{\mcitedefaultmidpunct}
{\mcitedefaultendpunct}{\mcitedefaultseppunct}\relax
\EndOfBibitem
\bibitem[Glass \latin{et~al.}(1995)Glass, Linde, and Negran]{glass1995high}
Glass,~A.; Linde,~D. v.~d.; Negran,~T. In \emph{Landmark papers on
  photorefractive nonlinear optics}; Pochi,~Y., Gu,~C., Eds.; World Scientific,
  1995; pp 371--373\relax
\mciteBstWouldAddEndPuncttrue
\mciteSetBstMidEndSepPunct{\mcitedefaultmidpunct}
{\mcitedefaultendpunct}{\mcitedefaultseppunct}\relax
\EndOfBibitem
\bibitem[Lalitha \latin{et~al.}(2007)Lalitha, Karazhanov, Ravindran,
  Senthilarasu, Sathyamoorthy, and Janabergenov]{lalitha2007electronic}
Lalitha,~S.; Karazhanov,~S.~Z.; Ravindran,~P.; Senthilarasu,~S.;
  Sathyamoorthy,~R.; Janabergenov,~J. {Electronic Structure, Structural and
  Optical Properties of Thermally Evaporated CdTe Thin Films}. \emph{Physica B}
  \textbf{2007}, \emph{387}, 227--238\relax
\mciteBstWouldAddEndPuncttrue
\mciteSetBstMidEndSepPunct{\mcitedefaultmidpunct}
{\mcitedefaultendpunct}{\mcitedefaultseppunct}\relax
\EndOfBibitem
\bibitem[Ma \latin{et~al.}(2021)Ma, Grushin, and Burch]{ma2021topology}
Ma,~Q.; Grushin,~A.~G.; Burch,~K.~S. {Topology and Geometry Under the Nonlinear
  Electromagnetic Spotlight}. \emph{Nat. Mater.} \textbf{2021}, \emph{20},
  1601--1614\relax
\mciteBstWouldAddEndPuncttrue
\mciteSetBstMidEndSepPunct{\mcitedefaultmidpunct}
{\mcitedefaultendpunct}{\mcitedefaultseppunct}\relax
\EndOfBibitem
\bibitem[Yang \latin{et~al.}(2017)Yang, Burch, and Ran]{yang2017divergent}
Yang,~X.; Burch,~K.; Ran,~Y. {Divergent Bulk Photovoltaic Effect in Weyl
  Semimetals}. \emph{arXiv} \textbf{2017}, 1712.09363,
  DOI:{\color{blue}{10.48550/arXiv.1712.09363}}\relax
\mciteBstWouldAddEndPuncttrue
\mciteSetBstMidEndSepPunct{\mcitedefaultmidpunct}
{\mcitedefaultendpunct}{\mcitedefaultseppunct}\relax
\EndOfBibitem
\bibitem[Chaubey and Van~Vliet(1986)Chaubey, and
  Van~Vliet]{chaubey1986transverse}
Chaubey,~M.~P.; Van~Vliet,~C.~M. {Transverse Magnetoconductivity of
  Quasi-Two-Dimensional Semiconductor Layers in the Presence of Phonon
  Scattering}. \emph{Phys. Rev. B} \textbf{1986}, \emph{33}, 5617\relax
\mciteBstWouldAddEndPuncttrue
\mciteSetBstMidEndSepPunct{\mcitedefaultmidpunct}
{\mcitedefaultendpunct}{\mcitedefaultseppunct}\relax
\EndOfBibitem
\bibitem[Togo and Tanaka(2015)Togo, and Tanaka]{togo2015first}
Togo,~A.; Tanaka,~I. {First Principles Phonon Calculations in Materials
  Science}. \emph{Scr. Mater.} \textbf{2015}, \emph{108}, 1--5\relax
\mciteBstWouldAddEndPuncttrue
\mciteSetBstMidEndSepPunct{\mcitedefaultmidpunct}
{\mcitedefaultendpunct}{\mcitedefaultseppunct}\relax
\EndOfBibitem
\bibitem[Carreras(2020)]{phonolammps}
Carreras,~A. {A Phonolammps: A Python Interface for LAMMPS Phonon Calculations
  Using Phonopy.} \textbf{2020}, \textit{Zenodo}
  {\color{blue}{https://doi.org/10.5281/zenodo.3940626}}\relax
\mciteBstWouldAddEndPuncttrue
\mciteSetBstMidEndSepPunct{\mcitedefaultmidpunct}
{\mcitedefaultendpunct}{\mcitedefaultseppunct}\relax
\EndOfBibitem
\bibitem[Kresse and Furthm{\"u}ller(1996)Kresse, and
  Furthm{\"u}ller]{kresse1996efficient}
Kresse,~G.; Furthm{\"u}ller,~J. {Efficient Iterative Schemes for Ab Initio
  Total-Energy Calculations Using a Plane-Wave Basis Set}. \emph{Phys. Rev. B}
  \textbf{1996}, \emph{54}, 11169\relax
\mciteBstWouldAddEndPuncttrue
\mciteSetBstMidEndSepPunct{\mcitedefaultmidpunct}
{\mcitedefaultendpunct}{\mcitedefaultseppunct}\relax
\EndOfBibitem
\bibitem[Kresse and Furthm{\"u}ller(1996)Kresse, and
  Furthm{\"u}ller]{kresse1996efficiency}
Kresse,~G.; Furthm{\"u}ller,~J. {Efficiency of Ab-Initio Total Energy
  Calculations for Metals and Semiconductors Using a Plane-Wave Basis Set}.
  \emph{Comput. Mater. Sci.} \textbf{1996}, \emph{6}, 15--50\relax
\mciteBstWouldAddEndPuncttrue
\mciteSetBstMidEndSepPunct{\mcitedefaultmidpunct}
{\mcitedefaultendpunct}{\mcitedefaultseppunct}\relax
\EndOfBibitem
\bibitem[Ernzerhof and Scuseria(1999)Ernzerhof, and
  Scuseria]{ernzerhof1999assessment}
Ernzerhof,~M.; Scuseria,~G.~E. {Assessment of the Perdew--Burke--Ernzerhof
  Exchange-Correlation Functional}. \emph{J. Chem. Phys.} \textbf{1999},
  \emph{110}, 5029--5036\relax
\mciteBstWouldAddEndPuncttrue
\mciteSetBstMidEndSepPunct{\mcitedefaultmidpunct}
{\mcitedefaultendpunct}{\mcitedefaultseppunct}\relax
\EndOfBibitem
\bibitem[Krukau \latin{et~al.}(2006)Krukau, Vydrov, Izmaylov, and
  Scuseria]{krukau2006influence}
Krukau,~A.~V.; Vydrov,~O.~A.; Izmaylov,~A.~F.; Scuseria,~G.~E. {Influence of
  the Exchange Screening Parameter on the Performance of Screened Hybrid
  Functionals}. \emph{J. Chem. Phys.} \textbf{2006}, \emph{125}, 224106\relax
\mciteBstWouldAddEndPuncttrue
\mciteSetBstMidEndSepPunct{\mcitedefaultmidpunct}
{\mcitedefaultendpunct}{\mcitedefaultseppunct}\relax
\EndOfBibitem
\bibitem[Mostofi \latin{et~al.}(2008)Mostofi, Yates, Lee, Souza, Vanderbilt,
  and Marzari]{mostofi2008wannier90}
Mostofi,~A.~A.; Yates,~J.~R.; Lee,~Y.-S.; Souza,~I.; Vanderbilt,~D.;
  Marzari,~N. {Wannier90: {{A}} Tool for Obtaining Maximally-Localised Wannier
  Functions}. \emph{Comput. Phys. Commun.} \textbf{2008}, \emph{178},
  685--699\relax
\mciteBstWouldAddEndPuncttrue
\mciteSetBstMidEndSepPunct{\mcitedefaultmidpunct}
{\mcitedefaultendpunct}{\mcitedefaultseppunct}\relax
\EndOfBibitem
\bibitem[Kim and Walsh(2021)Kim, and Walsh]{kim2021ab}
Kim,~S.; Walsh,~A. {Ab Initio Calculation of the Detailed Balance Limit to the
  Photovoltaic Efficiency of Single P-N Junction Kesterite Solar Cells}.
  \emph{Appl. Phys. Lett.} \textbf{2021}, \emph{118}, 243905\relax
\mciteBstWouldAddEndPuncttrue
\mciteSetBstMidEndSepPunct{\mcitedefaultmidpunct}
{\mcitedefaultendpunct}{\mcitedefaultseppunct}\relax
\EndOfBibitem
\end{mcitethebibliography}

\end{document}